\documentclass[aps,pre,superscriptaddress,twocolumn,nofootinbib,longbibliography]{revtex4-1}
\usepackage{graphicx}
\usepackage[normalem]{ulem}
\usepackage{amsmath,amssymb}
\usepackage{mathtools,mathrsfs,dsfont}
\usepackage{bm, physics}
\usepackage[dvipsnames]{xcolor}
\usepackage[colorlinks, linkcolor=Red, citecolor=NavyBlue, urlcolor=NavyBlue]{hyperref}
\usepackage[all]{hypcap} 

\renewcommand{\d}{\downarrow}
\renewcommand{\u}{\uparrow}

\begin{document}
\title{Domain coarsening in fractonic systems: a cascade of critical exponents}
\author{Jacopo Gliozzi}
\affiliation{Institute for Condensed Matter Theory, University of Illinois Urbana-Champaign, Urbana, USA}
\author{Federico Balducci}
\affiliation{Max Planck Institute for the Physics of Complex Systems, N\"othnitzer Str.\ 38, 01187 Dresden, Germany}
\author{Giuseppe De Tomasi}
\affiliation{CeFEMA-LaPMET, Departamento de F\'isica, Instituto Superior T\'ecnico, Universidade de Lisboa, Portugal}
\begin{abstract}
We study the dynamics of domain growth when multipole moments of the order parameter are conserved.
Following a quench into the ordered phase of the Ising model, the typical size of domains grows with time as 
$R(t) \sim t^{1/2}$ in the absence of conserved quantities. 
When the order parameter is conserved, the domain growth slows to $R(t) \sim t^{1/3}$. 
Conservation of higher moments of the order parameter fundamentally modifies this behavior: coarsening proceeds via anomalously slow growth. 
We analytically and numerically show that conservation of the $m$-th multipole moment causes domains to grow as $R(t) \sim t^{1/(2m+3)}$. This cascade of dynamical critical exponents characterizes a new family of non-equilibrium universality classes for fractonic systems.
\end{abstract}

\maketitle
\section{Introduction}
When a system is suddenly cooled from a disordered phase to an ordered phase, long-range order does not appear instantly. Instead, information about the direction of symmetry-breaking must spread through the system before the order parameter acquires a nonzero global value. This dynamical process, known as phase-ordering kinetics or coarsening, is central to non-equilibrium statistical mechanics~\cite{Bray1994review, Krapivsky}. 
An equivalent view focuses on the dynamics of topological
defects: global order is established once all defects are eliminated. 
If the defects are domain walls, coarsening corresponds to a growth in the typical linear size of domains, $R(t)$. 

In clean systems, the domain size generally grows as a power law, $R(t) \sim t^{1/z}$, where $z$ is the so-called coarsening exponent~\cite{Bray1994review}. 
At late times, the value of $z$ only depends on universal properties of the system, such as conservation laws and the type of order parameter. Very different microscopic models can therefore display the same large-scale behavior. 
However, a complete classification of dynamical universality classes out of equilibrium is still missing, as renormalization group methods are difficult to apply without time-translation invariance~\cite{Berges2009}. 
\begin{figure}[t!]
    \centering
    \includegraphics[width=\linewidth]{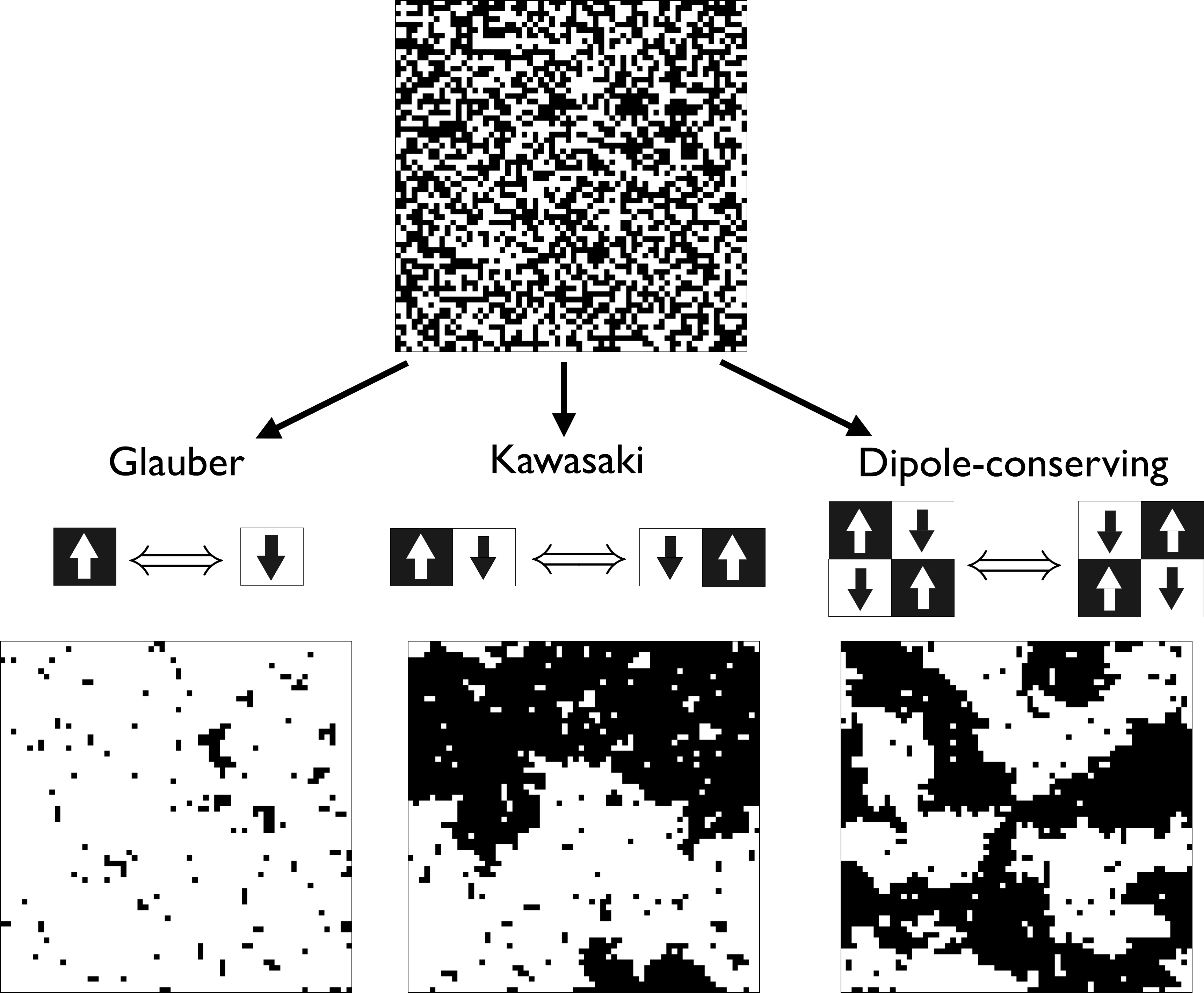}
    \caption{Spin dynamics in the kinetic Ising model following a quench from $T=\infty$ (top) to $T<T_c$. Late time snapshots of the domains are shown on the bottom for non-conserving (left), spin-conserving (middle), and spin-and-dipole-conserving (right) dynamics.}
    \label{fig:scheme}
\end{figure}

A well-studied universality class is that of Glauber dynamics, where an Ising-like order parameter is not conserved, as shown in Fig.~\ref{fig:scheme}, and $z=2$~\cite{Glauber1963}.
Here, domain growth is driven by the motion of domain walls, which evolve to locally reduce their curvature. 
If the order parameter is conserved, on the other hand, the domain growth is slowed to $z=3$. This universality class describes phase separation in binary alloys and spin-conserving Kawasaki dynamics in the Ising model~\cite{Kawasaki1966}, see Fig.~\ref{fig:scheme}. Due to the conservation law, domains walls cannot move independently. Instead, domains grow as the order parameter is redistributed in space, making the process diffusion-limited.

Glauber and Kawasaki dynamics demonstrate how conservation laws significantly influence domain growth, but there are clearly other possibilities. 
Recently, a new class of dynamical constraints has emerged from studies of fracton models~\cite{Nandkishore_2019,Vijay_2015,Vijay_2016,Pretko2017, Pretko_2018, Sala2020, Khemani2020, Gromov2020, Feldmeier2020}. Such systems conserve both a total U$(1)$ charge and its higher spatial moments, like dipole or quadrupole moments. 

The conservation of multipole moments fundamentally slows the hydrodynamics of charge transport. For example, a conserved charge normally diffuses, but
isolated charges cannot move if dipole moment is conserved. 
Instead, charge transport can only occur via the motion of dipoles, leading to subdiffusion~\cite{Gromov2020, Feldmeier2020, Ogunnaike_2023, Morningstar_2023, Gliozzi2023}.
Multipole constraints can also fragment the configuration space into exponentially many disconnected sectors, leading to glassy dynamics and ergodicity-breaking even at infinite temperature~\cite{Sala2020, Khemani2020}. However, the impact of multipole conservation on low-temperature phenomena such as coarsening remain largely unexplored. 

In this work, we investigate phase-ordering kinetics in the presence of multipole conservation laws.  We focus on systems whose microscopic dynamics conserves both a scalar order parameter and a subset of its higher moments; see Fig.~\ref{fig:scheme} for an example of the dipole-conserving case. 
These additional dynamical rules lead to anomalously slow domain coarsening.
By linking domain growth to the subdiffusive transport of the order parameter, we analytically show that $z = 2m+3$ in systems conserving up to the $m$-th multipole moment. In particular, if the magnetization and the center of mass of domains are both conserved ($m=1$), then $R(t) \sim t^{1/5}$. 

We support our theoretical predictions with numerical simulations of the two-dimensional Ising model. To reliably capture the asymptotic scaling behavior, we push our simulations to ten orders of magnitude in Monte Carlo steps, and further characterize the leading finite-time corrections to $R(t)$. In passing, we also prove that multipole conservation does not necessarily lead to frozen dynamics. Indeed, states with arbitrarily large domains are always dynamically accessible if the spatial range of local spin updates is large enough.

The plan of the paper is as follows. In Sec.~\ref{sec:model}, we define the model under consideration and introduce the main quantities used to characterize domain growth. In Sec.~\ref{sec:analytics}, we present our analytical results: first from a heuristic and intuitive perspective, and then more rigorously using a field-theory approach. 
These analytical predictions are supported by Monte Carlo simulations, presented in Sec.~\ref{sec:num_methods}. Fractonic systems exhibit strong mobility constraints, so it is not a priori obvious that domains can grow indefinitely. In Sec.~\ref{sec:proof}, we provide a simple combinatorial argument showing that one can indeed generate arbitrarily large domains through $m$-pole-conserving dynamics. Finally, in Sec.~\ref{sec:conclusion}, we summarize our conclusions and outline possible directions for future research.

\section{Model}
\label{sec:model}

We study the ordering of magnetic domains in the classical Ising model, defined by the Hamiltonian
\begin{equation}
\label{eq:ham}
    H = -\sum_{\langle \bm{x}, \bm{y} \rangle} \sigma_{\bm{x}} \sigma_{\bm{y}},
\end{equation}
where the sum runs over nearest-neighboring pairs of spins $\sigma_{\bm{x}} = \pm 1$. The spins are taken to live on a $d$-dimensional hypercubic lattice of linear size $L$, and periodic boundary conditions are employed. In dimension $d \geq 2$, this model hosts a phase transition between a high-temperature disordered phase and a low-temperature ordered phase. The Hamiltonian in Eq.~\eqref{eq:ham} determines the probabilities of different spin configurations through the Boltzmann distribution,
\begin{equation}\label{eq:boltzmann}
p_\text{eq}(\{\sigma_{\bm{x}}\}) \propto e^{-\beta H[\{\sigma_{\bm{x}}\}]},
\end{equation}
where $\beta = 1/T$ is the inverse temperature. At high temperatures, all configurations are approximately equally likely, and spins are randomly oriented. 
Below the critical temperature $T_c$, nearby spins align into contiguous domains. The phase transition is detected by a local order parameter, in this case the magnetization,  
\begin{equation} \label{eq:magnetization}
    M = \sum_{\bm{x}} \sigma_{\bm{x}},
\end{equation}
whose density is nonzero in the ordered phase ($T < T_c$) and vanishes in the disordered phase ($T > T_c$).

When a random spin configuration at $T=\infty$ is suddenly quenched to a temperature $T<T_c$, nearby spins begin to align and domains grow as the system relaxes toward a new equilibrium. 
This non-equilibrium process cannot be solely determined by the Hamiltonian in Eq.~\eqref{eq:ham}, which does not specify any intrinsic dynamics.
Instead, the system must be endowed with additional dynamical rules.
The properties of the resulting \emph{kinetic} Ising model depend crucially on the nature of these dynamics and, in particular, on the presence of conservation laws.

One possible choice is \emph{Glauber dynamics}~\cite{Glauber1963}, whereby randomly selected spins are flipped at a rate $e^{-\beta \Delta E}$, where $\Delta E$ is the change in energy due to the spin flip. This dynamics does not conserve the order parameter in Eq.~\ref{eq:magnetization}, and describes the formation of ferromagnetic domains. Another choice is \emph{Kawasaki dynamics}~\cite{Kawasaki1966}, in which nearest-neighbor spins are exchanged at a rate $e^{-\beta \Delta E}$, with $\Delta E$ the energy cost of the exchange. Here the order parameter remains conserved throughout the dynamics, which can therefore model phase separation in binary alloys or emulsions. 

Glauber and Kawasaki dynamics are commonly implemented as Monte Carlo algorithms that probabilistically update the spins of a system. In both cases, the spin update rates between configurations $\sigma$ and $\sigma'$ satisfy detailed balance:
\begin{equation}
    \label{eq:detailed}
    \frac{W(\sigma \rightarrow \sigma')}{W(\sigma' \rightarrow \sigma)} 
    = \frac{p_\text{eq}(\sigma')}{p_\text{eq}(\sigma)},
\end{equation}
where $W(\sigma \rightarrow \sigma')$ is the transition rate from $\sigma$ to $\sigma'$.
This condition guarantees that the steady state of the dynamics is the equilibrium Boltzmann distribution. Although both Glauber and Kawasaki dynamics reach the same equilibrium, the additional conservation law makes domains grow significantly slower under Kawasaki dynamics.

To quantify domain growth, we consider the equal-time spin-spin correlation function,
\begin{equation}
    \label{eq:corr}
    C(\bm{r}, t) = \expval{\sigma_{\bm{x} + \bm{r}}(t) \sigma_{\bm{x}}(t)},
\end{equation}
where the average is taken over different initial configurations, evolution histories, and positions $\bm{x}$. At long times after the initial quench, extensive experimental and numerical evidence suggests that the system is characterized by a single length scale $R(t)$~\cite{Lifshitz1961, Wagner1961,Binder1974, Marro1979, Furukawa1978, Furukawa1979, Furukawa1985, Coniglio1989}, loosely associated with the typical size of domains at time $t$.
This observation is known as the \emph{dynamical scaling hypothesis}, and it implies that the correlation function acquires the simple scaling form
\begin{equation}
    \label{eq:scaling}
    C(\bm{r}, t) = f \left(\frac{r}{R(t)}\right).
\end{equation}
In this scaling regime, $R(t)$ is much greater than the equilibrium correlation length $\xi$. Consequently, the system is a collection of equilibrated domains separated by thin domain walls that evolve over time. 

The dynamical scaling hypothesis holds in a variety of different systems~\cite{Bray1994review}. In models without frustration or quenched disorder~\cite{Shore1991, Cugliandolo2015}, such as ours, the growth of the typical domain size is usually power-law,
\begin{equation}
    \label{eq:dom_size}
    R(t) \sim t^{1/z}.
\end{equation}
Here, $z$ is the coarsening exponent, a critical exponent that depends on the non-equilibrium universality class of the dynamics. For local, non-conserving dynamics such as Glauber updates, $z=2$. For order parameter-conserving dynamics such as Kawasaki updates, the domain growth is slower, with $z=3$. 

In this work, we consider dynamics that conserve both the order parameter and its spatial moments, as in fractonic systems~\cite{Pretko2017, Gromov2020, Feldmeier2020}. Specifically, spin updates are taken to conserve up to the $m$-th multipole moment
\begin{equation}\label{eq:multipole_definition}
    P^{(m)}_{i_1, \ldots, i_m} = \sum_{\bm{x}} x_{i_1} \ldots x_{i_m} \sigma_{\bm{x}}.
\end{equation}
The $m=0$ moment is simply the total magnetization, corresponding to the usual Kawasaki dynamics. For $m \geq 1$, higher moments of the order parameter are also conserved, such as dipole ($m=1$) or quadrupole ($m=2$) moments. These additional conservation laws act as kinetic constraints that impede the motion of spins. The evolution of domains, which are formed by many individual spins, is also more constrained. For instance, Kawasaki dynamics preserves the total volume of spin-up domains, while dipole-conserving dynamics also conserves their center of mass (see Fig.~\ref{fig:scheme}).

In recent years, the out-of-equilibrium dynamics of multipole-conserving systems has been intensely studied, ranging from anomalous charge transport to ergodicity-breaking in quantum systems~\cite{Feldmeier2020, Sala2020, Khemani2020, Gromov2020, Gliozzi2023, Zechmann2023, Zechmann2024, Moudgalya2020, Dubinkin2021, Stahl2022, Morningstar2020, Pozderac_2023, Morningstar_2023, Ogunnaike_2023, Patil2023, Burnell_2024, Gliozzi_2024, Gliozzi2025}. However, most of these investigations have focused on the infinite-temperature regime, and comparatively little is known about low-temperature dynamics. At infinite temperature, all allowed transitions occur with equal probability, and the kinetic constraints alone determine the dynamics. 
At finite temperatures, energetic barriers emerge between configurations. These barriers can direct systems toward a restricted set of states, allowing for phenomena like metastability and domain growth.

Even at $T=\infty$, kinetic constraints fragment the configuration space into exponentially many disconnected sectors~\cite{Sala2020,Khemani2020, DeTomasi2019, Yang2020}. If spin update rules are too short-ranged, the dynamics is \emph{strongly fragmented}: the largest connected sector remains exponentially small compared to the full configuration space.\footnote{The largest sector contains $|\mathcal{C}_\text{largest}| \sim 2^{\alpha L^d}$ states, with $\alpha<1$. On the other hand, the number of allowed configurations at a fixed multipole moment $P^{(m)}$ is $|\mathcal{C}_\text{sym}|\sim 2^{L^d}$.} Practically, this implies that most configurations are dynamically inaccessible due to stringent conservation laws. For longer-ranged updates, the fragmentation is \emph{weak}, and the largest connected component approaches the full configuration space as $L\rightarrow \infty$. 

Strongly fragmented models are non-ergodic by definition, and domains likely cannot grow due to the absence of allowed transitions. 
We therefore focus on weakly fragmented dynamics, where the configuration space is still sufficiently connected to permit domain growth.  
Nevertheless, there is a competition between the multipole-conserving updates and the underlying Ising Hamiltonian. On one hand, thermal fluctuations favor the growth of ordered domains at low temperatures. 
On the other hand, the kinetic constraints imposed by the updates hinder spin transport, making it difficult to form large domains.
Whether this competition leads to frozen dynamics at low temperatures~\cite{Morningstar2020, Pozderac_2023, Cornell1991, Ben-Naim1998} is a non-trivial question that we postpone to Sec.~\ref{sec:proof}. 
In what follows, we assume that arbitrarily large domains can still form, and provide analytical and numerical evidence for the resulting coarsening exponents.

\section{Domain growth exponents in presence of multipole conservation}
\label{sec:analytics}

The critical exponents that govern phase ordering depend only on large-scale properties of the system, like the nature of the order parameter and the symmetries of the dynamics. In the Ising model, the order parameter is a scalar that encodes the local magnetization. The topological defects of this order parameter are codimension-one domain walls. Following a quench into the ordered phase, these defects move and annihilate to lower the overall free energy. Their evolution, constrained by the symmetries of the system, fully characterizes the growth of the late-time domains.

In keeping with these general considerations, we derive the domain coarsening exponents using a minimal model of multipole-conserving dynamics with a scalar order parameter. First, in Sec.~\ref{sec:subdiffusion} we review how multipole symmetry slows the transport of the order parameter. Based on this transport, we provide a heuristic argument in a 1D toy model (Sec.~\ref{sec:1d}) and develop a continuum model for $d > 2$ (Sec.~\ref{sec:continuum}). Using the continuum model, we extract the domain growth law both with a scaling argument (Sec.~\ref{sec:scaling_argument}) and based on the relaxation of a perturbed domain wall (Sec.~\ref{sec:defect_relax}). We also compute the early-time corrections to the asymptotic scaling in Sec.~\ref{sec:early_time}, as these become particularly severe in multipole-conserving models and are needed to interpret numerical results. Notice that we make use of a continuous time model, which inherently assumes that the final results do not depend qualitatively on the specific moves being performed in a discrete-time, numerical setting. 

\subsection{Subdiffusion of the order parameter}
\label{sec:subdiffusion}
In this section, we assume that the order parameter (and eventually its moments) is conserved by the dynamics, and discuss how it spreads throughout the system. We will refer to the order parameter as ``charge'' to stress the connection with transport phenomena.

The conservation of a scalar local charge $\phi$ is described by the continuity equation:
\begin{equation}
    \label{eq:continuity_charge}
    \partial_t \phi + \partial_i J_i = 0,
\end{equation}
where $J_i$ is the current of the order parameter in the $i$-direction, and summation over repeated indices is implied. The hydrodynamics of charge follows from writing $J_i$ as a derivative expansion in $\phi$ and keeping only lowest-order terms that obey the symmetry. The simplest expression is Fick's law:
\begin{equation}\label{eq:fick}
    J_i \propto -\partial_i \phi,
\end{equation}
where the negative sign reflects flow from high to low density.
Inserting Eq.~\eqref{eq:fick} into Eq.~\eqref{eq:continuity_charge} gives the diffusion equation:
\begin{equation}
    \label{eq:diff}
    \partial_t \phi = D \nabla^2 \phi,
\end{equation}
where $D$ is the diffusion constant.
From the anisotropic scaling of space and time, it is clear that the conserved charge diffuses a distance $\mathcal{O}(t^{1/2})$ in time $t$.

In a system that conserves both the total U$(1)$ charge $\int \phi(x) d^dx$ and its dipole moment, $P^{(1)}_i = \int x_i \phi(x) d^d x$ (where we have written Eq.~\eqref{eq:multipole_definition} in the continuum), the continuity equation still holds.  
Microscopically, however, individual charges cannot move because of the dipole constraint. Charge can only evolve through the hopping of dipoles, so the charge current must be expressed in terms of a more fundamental dipole current~\footnote{{In a (quantum) lattice model, the charge and dipole currents can be discretized as $J_\text{charge} \sim -i (\sigma^+_r \sigma^-_{r+1} - \text{h.c.})$ and $J_\text{dip} \sim -i (\sigma^+_r \sigma^-_{r+1}\sigma^-_{r+2} \sigma^+_{r+3} - \text{h.c.})$, where $\sigma^{\pm}$ flips a spin up/down.}}. The two currents are related by
\begin{equation}
    \label{eq:charge_dip_curr}
    J_{i} = \partial_j J^\text{dip}_{ij},
\end{equation}
where $J^\text{dip}_{ij}$ is the current of $j$-oriented dipoles moving in the $i$-direction~\cite{Gromov2020}. Heuristically, an inhomogeneous flow of dipoles generates a net flow of charge. 

Next, $J^\text{dip}_{ij}$ must be related to the charge density through a hydrodynamic assumption. Keeping the lowest order derivatives of $\phi$ consistent with dipole symmetry,
\begin{equation}
    \label{eq:dip_curr}
    J^\text{dip}_{ij} \propto \partial_i \partial_j \phi.
\end{equation}
Roughly speaking, the two derivatives are necessary because both charge and dipole moment are conserved~\cite{Gromov2020, Feldmeier2020}. More precisely, a system with open boundary conditions requires $J^\text{dip}_{ij} = 0$ in equilibrium. Inserting Eq.~\eqref{eq:dip_curr} into this condition yields the equilibrium configuration $\phi(x_i) = a + b_i x_i$. The resulting $(d+1)$ free parameters exactly match the number of global conserved quantities needed to characterize equilibrium: the total charge and its $d$ dipole moments.

Combining the continuity equation with Eqs.~\eqref{eq:charge_dip_curr} and \eqref{eq:dip_curr}, one arrives at the subdiffusion equation,
\begin{equation}
    \label{eq:subdiff} 
    \partial_t \phi = -D \nabla^4 \phi,
\end{equation}
where $D$ is now a subdiffusion constant. The scaling of space and time shows that the resulting charge transport is slower than diffusive: charge only travels an $\mathcal{O}(t^{1/4})$ distance in time $t$. If up to $m$-th moments of the charge are conserved, the relevant subdiffusion equation becomes
\begin{equation}
    \label{eq:subdiff_m} 
    \partial_t \phi = -D (-\nabla^2)^{m+1} \phi.
\end{equation}

\subsection{One-dimensional example}
\label{sec:1d}

To understand how charge transport affects the motion of domain walls, it is helpful to review Glauber and Kawasaki dynamics in one dimension. Although the 1D Ising model does not have a finite-temperature phase transition, domain growth still occurs after a quench to $T \approx 0$. The characteristic time scale of coarsening can be estimated by using the typical time needed to flip an entire domain~\cite{Cardy}.

In Glauber dynamics, a single spin flip either creates domains walls ($\u\u\u \Rightarrow \u\d\u$), annihilates domain walls ($\u\d\u \Rightarrow \u\u\u$), or translates a domain wall ($\u\d\d \Rightarrow \u\u\d$). 
At low temperatures, creating domain walls is energetically costly, so few are produced. At late times, domain walls are widely separated and annihilation events are likewise rare. Domain growth is therefore effectively governed by a random walk of domain walls. Domain walls therefore move diffusively, and the characteristic time to eliminate a domain of length $R$ scales as $t \sim R^2$, implying a dynamical exponent $z=2$.

In Kawasaki dynamics, isolated domain walls cannot move because the total magnetization is conserved. Instead, domains grow via spin exchange: the spins themselves execute a random walk, leading to coarsening. A single down spin travels through an up-spin domain of length $R$ in a time $\mathcal{O}(R^2)$. However, this process only shifts the domain by one lattice spacing. To fully flip the domain, $\mathcal{O}(R)$ down spins must traverse it, leading to a characteristic time scale of $t \sim R^3$ and $z=3$.

The same argument can be heuristically applied to multipole-conserving systems. Once again, the total spin is conserved, so $\mathcal{O}(R)$ opposite spins must move through a domain of length $R$ to flip it. However, the spins no longer diffuse because of the additional conservation laws. For example, in dipole-conserving dynamics, the time for a spin to subdiffuse over a distance $R$ scales like $R^4$, giving a characteristic time of $t\sim R^5$. Using the charge transport in Eq.~\eqref{eq:subdiff_m}, this approach predicts a critical exponent $z = 5$ for dipole conservation, and generically $z=2m + 3$ for $m$-pole conserving systems.

The above argument for the coarsening exponent is merely suggestive and obscures a number of subtleties in the one-dimensional case. First, the coarsening regime requires a delicate balance of low but nonzero temperature to avoid infinitely long-lived metastable states~\cite{Ben-Naim1998}. Moreover, domain wall curvature is completely absent in one dimension, but plays a crucial role in domain growth for $d\geq 2$. Lastly, the transport of a single spin across an opposite domain, even if subdiffusive, explicitly violates dipole-moment conservation. Indeed, down spins can only hop in the vicinity of other down spins, so motion through bulk domains is restricted. Despite these caveats, we now show that this heuristic argument correctly predicts the dynamical exponent $z=2m+3$ for multipole-conserving dynamics in arbitrary spatial dimension.

\subsection{Continuum model}
\label{sec:continuum}

The equilibrium physics of a model with a scalar order parameter $\phi$ can be described by a phenomenological Landau-Ginzburg free energy:
\begin{equation}
    \label{eq:landau}
    F[\phi] = \int d^d x \left[\frac{1}{2}(\nabla \phi)^2 + V(\phi)\right],
\end{equation}
where $V(\phi)$ is an interaction potential. In the ordered phase of the Ising model, $V(\phi) = (1 - \phi^2)^2$ has a double-well structure, with its two minima representing the two symmetry-breaking equilibrium states. The gradient term provides an energy penalty for misaligned spins and is responsible for the surface tension of domain walls. 

After a quench into the ordered phase, the system evolves to the equilibrium set by Eq.~\eqref{eq:landau}, but the nature of the evolution depends on the conserved quantities.
When the order parameter is not conserved, the simplest model for dynamics is the time-dependent Landau-Ginzburg (TDLG) equation: 
\begin{equation}
    \label{eq:tdlg}
    \partial_t \phi = - \frac{\delta F}{\delta \phi} = \nabla^2 \phi - V'(\phi),
\end{equation}
where an overall constant of proportionality has been absorbed into the time scale. 
The TDLG equation, or model A in the Hohenberg-Halperin-Wagner nomenclature~\cite{Hohenberg1977}, represents a dissipative evolution of the order parameter towards a minimum in the free energy.\footnote{At finite temperature, the dynamics should be supplemented by a noise term to model thermal fluctuations. However, as temperature is an irrelevant parameter (in the RG sense) for coarsening, one can ignore this detail~\cite{Bray1994review}.} 

If the order parameter is conserved, on the other hand, its time derivative must obey the continuity equation Eq.~\eqref{eq:continuity_charge}. The current $J_i$ should depend on $\phi$ through the free energy, and the simplest choice that flows to a free energy minimum is
\begin{equation}
    \label{eq:fick_generalized}
    J_i \propto -\partial_i \left(\frac{\delta F}{\delta \phi}\right).
\end{equation}
Identifying $\mu = \delta F / \delta \phi$ as the chemical potential associated with the order parameter, this relation is simply the generalization of Fick's law. For small fluctuations around equilibrium, $\mu \sim \phi$, and Eq.~\eqref{eq:fick_generalized} reduces to Eq.~\eqref{eq:fick}. In the general case, the order parameter flows across a chemical potential gradient, and one arrives at the Cahn-Hilliard equation (model B)~\cite{Cahn1958}:
\begin{equation}
    \label{eq:ch}
    \partial_t \phi = \nabla^2 \left(\frac{\delta F}{\delta \phi}\right) = - \nabla^2 [\nabla^2 \phi - V'(\phi)].
\end{equation}

When up to $m$-th multipole moments of the order parameter are conserved, the Cahn-Hilliard equation can be straightforwardly generalized. Repeating the same arguments that led to the subdiffusion equation in Eq.~\eqref{eq:subdiff_m}, the order parameter evolves as
\begin{equation}
     \label{eq:ch_dipole}
    \partial_t \phi = -(-\nabla^2)^{m+1} \left(\frac{\delta F}{\delta \phi}\right) = (-\nabla^{2})^{m+1} [\nabla^2 \phi - V'(\phi)].
\end{equation}
In the dipole-conserving case, for example, the dipole current is taken to be 
\begin{equation}
    \label{jdip_landau}
    J^\text{dip}_{ij} \propto \partial_i \partial_j \left(\frac{\delta F}{\delta \phi}\right),
\end{equation}
to obtain Eq.~\eqref{eq:ch_dipole} for $m=1$.
Using this \emph{multipolar} Cahn-Hilliard equation as the starting point, one can derive the growth laws for domains evolving under a multipole constraint.

\subsection{Scaling argument for domain growth}
\label{sec:scaling_argument}
Late-time coarsening is fundamentally driven by the surface tension of domain walls. Highly corrugated interfaces cost more energy, and must therefore be straightened to reach equilibrium. Without conserved quantities, domain walls can move freely to reduce their curvature. Moreover, this evolution is local, as different regions of an interface evolve independently. On the other hand, conserved quantities impose global constraints on the motion of interfaces. Although domain walls must still smoothen to dissipate energy, this process can only happen through a global rearrangement of the conserved charge.

To obtain the coarsening exponents, we review Huse's scaling argument for the charge-conserving case~\cite{Huse1986} and then generalize it to multipole conservation. An important ingredient in this approach is the chemical potential near a domain wall, which is given by the Gibbs-Thomson relation:
\begin{equation}
    \label{eq:gibbsthomson}
    \mu = -\frac{\sigma \kappa}{\Delta \phi},
\end{equation}
where $\sigma$ is the surface tension of the interface, $\kappa$ is its curvature, and $\Delta \phi$ is the change in order parameter across it. This relation is a standard consequence of the Landau-Ginzburg free energy, and for convenience we derive it in App.~\ref{app:gibbsthomson}.

As a warm-up, let us first revisit the non-conserving case. Plugging the Gibbs-Thomson relation into the TDGL equation, $\partial_t \phi = - \frac{\delta F}{\delta \phi} \sim \mu$, one finds $\partial_t \phi \sim -\kappa$ near an interface of curvature $\kappa$. This curvature-driven flow of the order parameter is known as the Allen-Cahn equation~\cite{Allen1972}. Here, the local velocity of the domain wall (interface) is simply given by $v_\text{int} = \partial_t \phi$. One can now leverage the dynamical scaling hypothesis, which states that the typical domain size, $R(t)$, is the only length scale present. As a result, the interface curvature is given by $\kappa \sim 1/R$ and its velocity by $v_{\text{int}} \sim \partial_t R$, leading to 
\begin{equation}
    \label{eq:allencahn}
    \partial_t R \sim \frac{1}{R}.
\end{equation}
Solving this equation yields the $R(t)\sim t^{1/2}$ scaling law of non-conserving (Glauber) dynamics.

When the order parameter is conserved, interfaces can advance only via transport of $\phi$ through the domains. Consequently, the normal velocity of an interface is set by the net current crossing it, $v_{\text{int}}\sim J$. Because the order parameter is transported diffusively, the current is proportional to the gradient of the chemical potential, $J\sim \nabla\mu$. For a domain of size $R$, the chemical potential near its interface scales as $\mu\sim 1/R$ by the Gibbs–Thomson relation and varies over distances of order $R$, so $|\nabla\mu|\sim \mu/R\sim 1/R^{2}$. The resulting current, $J\sim 1/R^{2}$, drives an overall domain growth
\begin{equation}
    \label{eq:one_third}
    \partial_t R \sim \frac{1}{R^{2}}.
\end{equation}
Solving this gives the scaling law $R(t)\sim t^{1/3}$ characteristic of order-parameter–conserving (Kawasaki) dynamics.

This argument can be extended to dynamics with dipole and higher-moment conservation. For instance, dipole conservation implies that the ordinary current is a derivative of dipole current, which itself scales as $J^\text{dip}_{ij} \sim \partial_i \partial_j \mu$. The domain wall velocity is then proportional to $J \sim \nabla(\nabla^2 \mu)$, which by the dynamical scaling hypothesis is $\mathcal{O}(1/R^4)$. The two additional derivatives translate to slower domain growth,
\begin{equation}
    \label{eq:one_fifth}
    \partial_t R \sim \frac{1}{R^4}.
\end{equation}
The size of domains therefore grows as $R\sim t^{1/5}$ under dipole-conserving dynamics. Applying the same argument for an $m$-pole conserving system, we obtain $z = 2m + 3$.

\subsection{Domain growth from defect relaxation}
\label{sec:defect_relax}

The simple scaling argument above provides an intuitive picture of how domains grow when subject to a multipole constraint. In this section, we also provide a slightly more rigorous derivation of the coarsening exponent. We follow Ref.~\cite{Bray1998perturb} in analyzing how domain walls relax after they are perturbed. Specifically, we consider a periodic modulation of an interface with wavevector $k$, and study its relaxation rate $\omega(k)$. The central dogma of the scaling hypothesis posits that a single length scale $R(t)$ governs this process. Taking $\omega \sim 1/t$ and $k\sim 1/R(t)$, the coarsening exponent can then be inferred from the relation $\omega \sim k^{z}$.

First, consider a static, flat domain wall in equilibrium. Denoting the direction normal to the wall by $r$, the profile of the order parameter $\phi_0(r)$ obeys the equilibrium Landau-Ginzburg equation of motion,
\begin{equation}
    \label{eq:eom}
    \partial_r^2 \phi_0 = V'(\phi_0).
\end{equation}
The domain wall has boundary conditions $\phi_0(r=\pm \infty) = \pm 1$, corresponding to the global minima of the double well potential $V(\phi) = (1-\phi^2)^2$. For simplicity, we also set $\phi_0(0) = 0$. To generate dynamics, a small oscillatory perturbation can be added to the interface, leaving an overall profile of
\begin{equation}
    \label{eq:phi_perturb}
    \phi(r, x_i, t) = \phi_0(r) + A \phi_1(r) e^{i \bm{k} \cdot \bm{x} - \omega t},
\end{equation}
where $A \ll 1$, and $x_i$ represent the directions tangent to the domain wall. The frequency of the perturbation is imaginary because the domain wall evolves dissipatively, lowering its energy to return to equilibrium.

To extract the relation $\omega\sim k^{z}$, one substitutes the perturbed domain wall into the multipolar Cahn-Hilliard equation, $ \partial_t \phi = -(-\nabla^2)^{m+1} \left(\frac{\delta F}{\delta \phi}\right)$, to model its relaxation dynamics. Linearizing in the perturbation, we find
\begin{equation}
    \label{eq:linearized_ch}
    [\partial_r^2 + V''(\phi_0)] \phi_1(r) + k^2 \phi_1(r) - \omega \int dr' G_k (r-r') \phi_1(r')  = 0,
\end{equation}
where the differential operator $[k^{2(m+1)} + (-1)^m \partial_r^{2(m+1)}]$ was inverted using its Green function\footnote{This operator arises from solving $(-\nabla^2)^{m+1} \psi(r) e^{i k \cdot x - \omega t} = \phi_1(r)e^{i k \cdot x - \omega t}$.}
\begin{equation}
    \label{eq:green}
    G_k(r - r') = \int \frac{dq}{2\pi} \frac{e^{i q (r-r')}}{q^{2(m+1)} + k^{2(m+1)}}.
\end{equation}
Let us examine Eq.~\eqref{eq:linearized_ch}, which describes how perturbations melt, and recall that our goal is to determine the dispersion relation $\omega = \omega(k)$. 
When $k=0$, the domain wall is merely shifted and does not relax, leaving $\omega=0$. This translational zero mode of the domain obeys the eigenvalue equation
\begin{equation}
    \label{eq:zeromode}
    [\partial_r^2 + V''(\phi_0)] \phi_1(r) = 0,
\end{equation}
which sets $\phi_1(r) = \partial_r \phi_0(r)$. 

We are interested in the relaxation rate for small but nonzero $k$, which captures the large-scale coarsening dynamics of the domain. Taking an inner product of both sides of Eq.~\eqref{eq:linearized_ch} with the eigenfunction $\phi_0'(r)$, we obtain
\begin{equation}
    \label{eq:fraction}
    \omega  = \frac{k^2 \int dr \phi_0'(r)^2}{\int dr \int dr' G_k(r-r') \phi'_0(r) \phi_0'(r')}.
\end{equation}
The integral in the numerator is a constant proportional to the surface tension of the interface, $\sigma = \int d^dx (\nabla \phi)^2$. To evaluate the denominator, note that $\phi_0(r)$ is a step-like function and therefore $\phi_0'(r)$ is only nonzero very close to the domain wall. As a result, it acts approximately as a delta function, leaving $G_k(0)$ in the denominator. From the definition of the Green function, $G_k(0)\sim 1/k^{2m+1}$, giving an overall scaling of 
\begin{equation}
    \omega \sim k^{2m+3}.
\end{equation}
The resulting domain wall scaling, $R(t) \sim t^{1/(2m+3)}$, exactly reproduces the results of the scaling argument.

As a result, the domain growth in multipole-conserving systems is markedly slower than in ordinary Glauber or Kawasaki dynamics. However, the fundamental mechanism is the same: domains walls move to reduce their curvature, and this motion is driven by the flow of spins. With multipole conservation, the key additional ingredient is that local spins cannot move on their own, leading to subdiffusion instead of diffusion of the order parameter.

\subsection{Early-time corrections to scaling}
\label{sec:early_time}
Above, we obtained the critical exponents that control late-time coarsening. At early times, however, activated processes that raise the energy are suppressed, leading to corrections to the domain growth law. These corrections are especially important in multipole conserving systems, where the ultraslow dynamics remains in the early time regime for longer.
\begin{figure}[t!]
    \centering
    \includegraphics[width=0.7\linewidth]{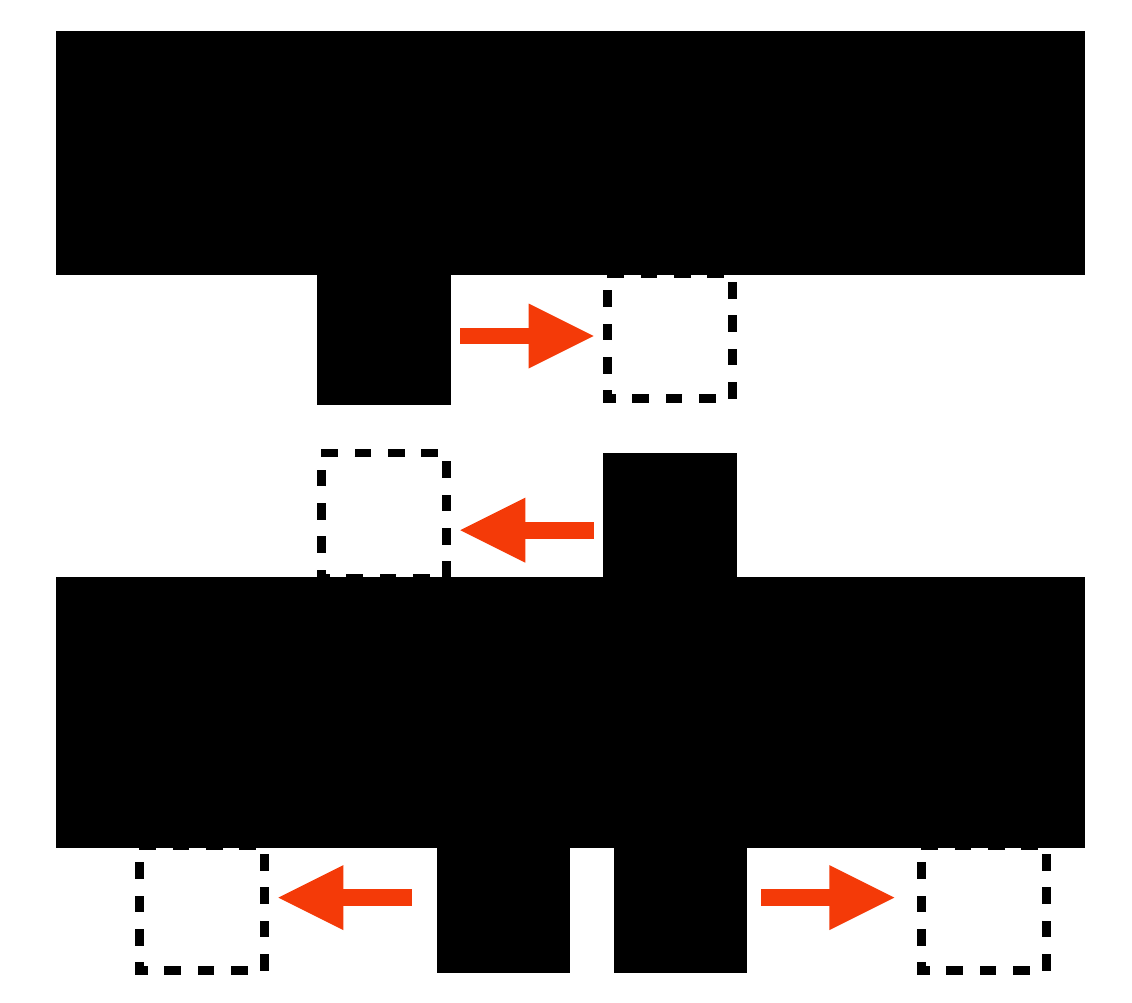}
    \caption{Two examples of spin updates that conserve dipole moment and also do not raise the overall energy. Spin evolution with $\Delta E = 0$ is confined to the surfaces of domains. 
    }
    \label{fig:surface}
\end{figure}

First, consider the case of Kawasaki dynamics. At times earlier than $e^{\beta \Delta E}$, where $\Delta E$ is the lowest positive energy spin update, the system can only evolve via energy-lowering and energy-neutral events. Such dynamics cannot create new domain walls, and thus spins travel only along existing ones. Since the early-time diffusion of spins occurs only on interfaces, the diffusion constant is renormalized by the interface density, $1/R$~\cite{Huse1986, vanGemmert2005}. The resulting domain size satisfies 
\begin{equation}
    \label{eq:kawasaki_correction}
    \partial_t R \sim \frac{1}{R^3},
\end{equation}
yielding a slower $t^{1/4}$ growth law. 

The origin of this scaling is a fundamentally different mechanism for domain growth. Bulk diffusion permits small domain bubbles to shrink and eventually vanish as they shed spins into the opposite domain surrounding them. At the end of this process, a small domain has vanished and a large one has grown, increasing the average domain size. Surface diffusion, in contrast, cannot transport spins from one domain to another. Nevertheless, a domain can still move by sliding spins along its surface. Eventually, two domains can meet each other and coalesce, also increasing the average domain size. The latter process is slower, and eventually yields to bulk diffusion at times much later than $e^{\beta \Delta E}$.

A similar early-time correction should arise with multipole conservation. For concreteness, consider a dipole-conserving update, which can be viewed as a coordinated hopping of two like spins in opposite directions. At early times, processes that create domain walls are energetically suppressed. Once again, spin transport is restricted to interfaces, but now both participating spins must be confined to an interface, as in Fig.~\ref{fig:surface}.
The probability of finding two nearby spins that can hop, each on a domain wall, scales as $1/R^2$. After rescaling the kinetic coefficient in Eq.~\eqref{eq:one_fifth}, the typical domain size evolves as
\begin{equation}
    \label{eq:dipole_correction}
    \partial_t R \sim \frac{1}{R^6}.
\end{equation}
As a result, domains grow as $t^{1/7}$ at early times, before eventually giving way to the late-time $z = 5$ behavior. Here the early-time correction is stronger than in the charge-conserving case, and effectively leads to the same coarsening exponent as late-time quadrupole-conserving dynamics. 

If higher moments of the order parameter are conserved, early-time corrections due to surface transport become even more severe. For example, a quadrupole-conserving update is akin to two simultaneous dipole-conserving updates, and therefore requires four nearby spins that can hop, each on a domain wall. The resulting kinetic coefficient is rescaled by $1/R^4$, yielding a growth law of $R(t) \sim t^{11}$. In general, the early-time correction to scaling in an $m$-pole conserving system gives an apparent coarsening exponent of $z =2m + 3 + 2^m$. The domain growth in such systems effectively frozen for all practical purposes until energy-raising events can occur.

Unlike Kawasaki dynamics, the multipole conserving case also features several intermediate regimes between ``early-time'' and ``late-time'' coarsening. These intermediate periods occur when only a subset of the possible energy-raising events are allowed. 
Consider dipole-conserving events, which hop two like spins. At early times, both spins are restricted to interfaces, and $R(t) \sim t^{1/7}$. However, if only one of the two spins is restricted to an interface, the kinetic coefficient is merely renormalized by $1/R$, giving $R(t)\sim t^{1/6}$. Such events cost energy $\Delta E_1$, which is lower than the energy required for both spins to hop into a domain $\Delta E_2$. At intermediate times $e^{\beta \Delta E_1} \ll t \ll e^{\beta \Delta E_2}$, the apparent coarsening exponent is therefore $z=6$, and similar results hold for higher multipoles.

\section{Simulations}
\label{sec:num}

In this section, we confirm our analytical prediction for the coarsening exponent via Monte Carlo simulations. First, in Sec.~\ref{sec:num_methods}, we explain our numerical procedures; then, in Sec.~\ref{sec:num_results}, we show the data obtained.

\begin{figure}[t!]
    \centering
    \includegraphics[width=0.8\linewidth]{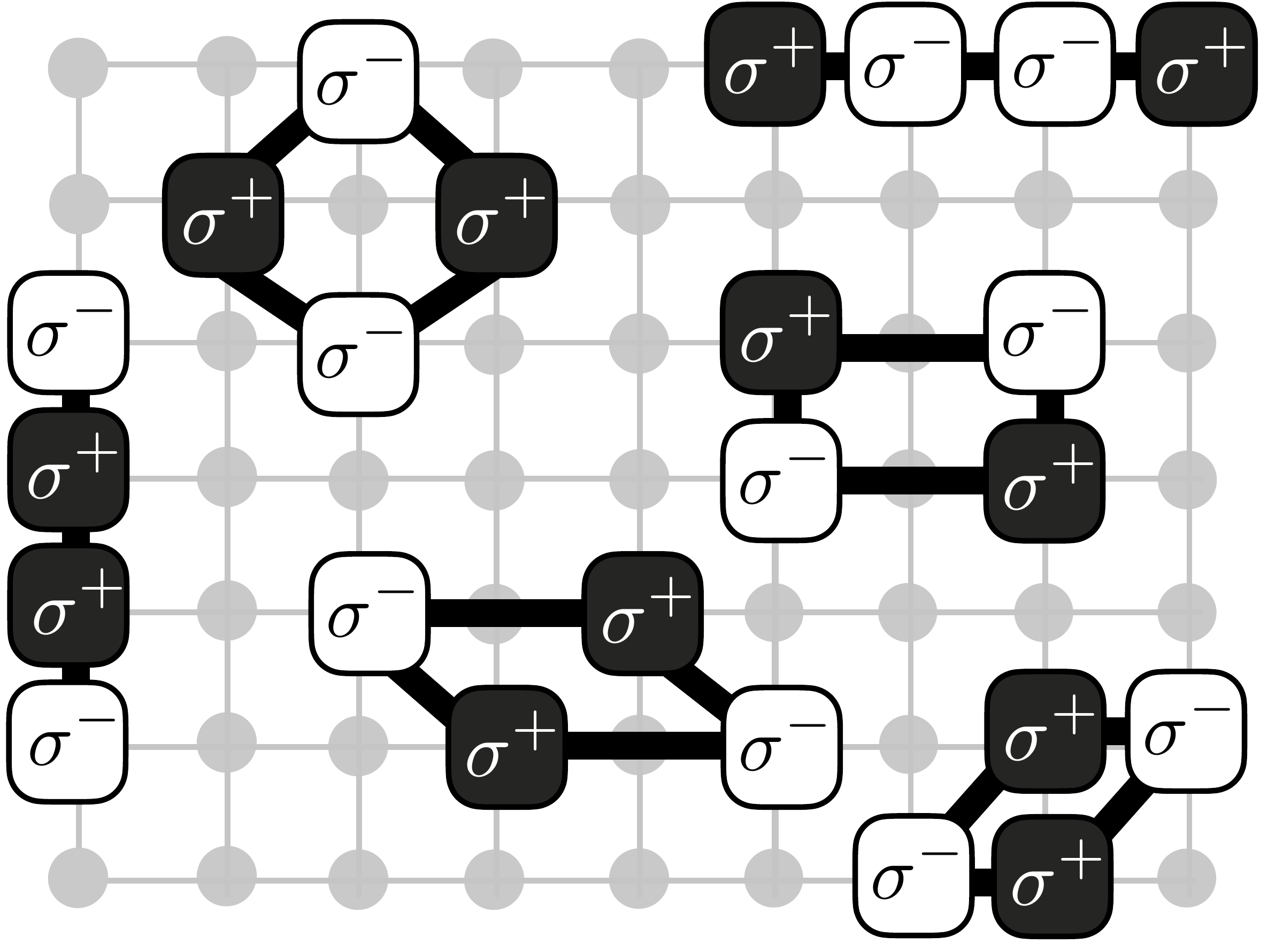}
    \caption{Examples of the local update rules used in dipole-conserving dynamics. Each update exchanges two dipoles whose length and separation are bounded by two lattice units to preserve locality.}
    \label{fig:gates}
\end{figure}

\subsection{Numerical methods}
\label{sec:num_methods}

We focus on the ferromagnetic Ising model on a two-dimensional square lattice of linear size $L$, with the Hamiltonian given by Eq.~\eqref{eq:ham}. The dynamical update rules of our Monte Carlo algorithm respect locality and the dipole (or multipole) constraint. To conserve the $m$-th multipole moment, an update must exchange two identically shaped $m$-poles. Each update therefore requires an initial site and $m+1$ vectors: $m$ vectors to specify the shape of the $m$-poles that are exchanged, and one for their separation. We use $\sigma^+$ ($\sigma^-$) to represent an update flipping a down (up) spin to an up (down) spin.\footnote{This notation is borrowed from the Pauli matrices commonly used in quantum spin-1/2 systems.} 

A generic dipole-conserving update can be written as
\begin{equation}
    \label{eq:dipole_gate}
    \sigma^+(\bm{r_0}) \sigma^-(\bm{r_0} + \bm{r_1}) \sigma^-(\bm{r_0} + \bm{r_2}) \sigma^+(\bm{r_0} + \bm{r_1} +\bm{r_2} ),
\end{equation}
where $\bm{r_0}$ is the position of the update, $\bm{r_1}$ specifies the size of the dipoles involved, and $\bm{r_2}$ specifies their separation. 
This process is the analogue of the charge-conserving update $\sigma^+(\bm{r}_0) \sigma^-(\bm{r_0} + \bm{r}_1)$, which exchanges a positive and negative charge. {Much like the dipole exchange of Eq.~\eqref{eq:dipole_gate}, a quadrupole-conserving updates corresponds to two simultaneous dipole exchanges.} In this way, an $m$-pole conserving update can be hierarchically constructed from lower moments~\cite{Feldmeier2020,Gliozzi2023}.
Examples of dipole-conserving updates are shown in Fig.~\ref{fig:gates}. Locality is enforced by taking vectors of magnitude no larger than two lattice units for dipole updates, or four lattice units for quadrupole updates. Furthermore, the periodic boundary conditions imply that multipole moments are only conserved modulo $L$.

\begin{figure}[t!]
    \centering
    \hspace{-0.4cm}
    \includegraphics[width=\linewidth]{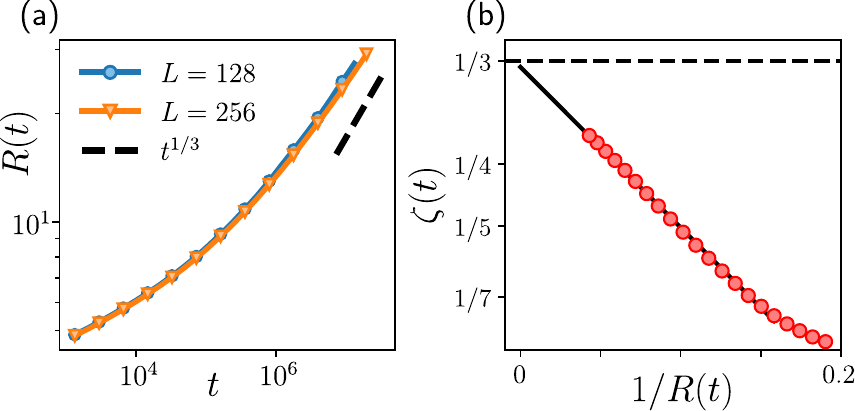}
    \caption{Domain growth with Kawasaki dynamics. (a) Long times are needed to access the asymptotic scaling $R(t) \sim t^{1/3}$. (b) Early-time corrections are visible from the running logarithmic derivative of $R(t)$, whose linear extrapolation to $R\rightarrow \infty$ is compatible with $z=3$ (horizontal dashed line).}
    \label{fig:Kawasaki}
\end{figure}

We always start from a random state with $P^{(m)}_{i_1\ldots i_m}=0$, which is equivalent to working at infinite temperature. The system is then quenched into the ordered phase by setting $T = 0.75\, T_c$, with $T_c = 2/\ln(1+\sqrt{2})$. At each Monte Carlo step, a random $m$-pole exchange is attempted. If the existing spin configuration supports the update, it is applied with probability $\text{min}(e^{\beta \Delta E}, 1)$, where $\Delta E$ is the resulting change in energy. This update probability is chosen to respect the detailed balance condition, Eq.~\eqref{eq:detailed}. The unit of time is $L^2$ Monte Carlo steps, i.e., one attempted exchange per spin.

Throughout the dynamics, we quantify the growth of domains using the equal-time correlation function $C({\bm r},t)$ in Eq.~\eqref{eq:corr}. At early times, the $C(\bm{r}, t)$ rapidly decays on length scales of order $\xi$, the equilibrium correlation length.
At late times, domains are significantly larger than $\xi$, and $C(\bm{r}, t)$ decays more slowly.\footnote{At $T = 0.75 \, T_c$, $\xi$ is approximately one lattice spacing.} 
In this coarsening regime, the correlation function approximately takes the scaling form of Eq.~\eqref{eq:scaling}, with a single length scale $R(t)$ quantifying domain size. This length scale can then be estimated using, e.g., the first zero of $C(\bm{r}, t)$~\cite{Huse1986}. 

\begin{figure}[t!]
    \centering
    \includegraphics[width=\linewidth]{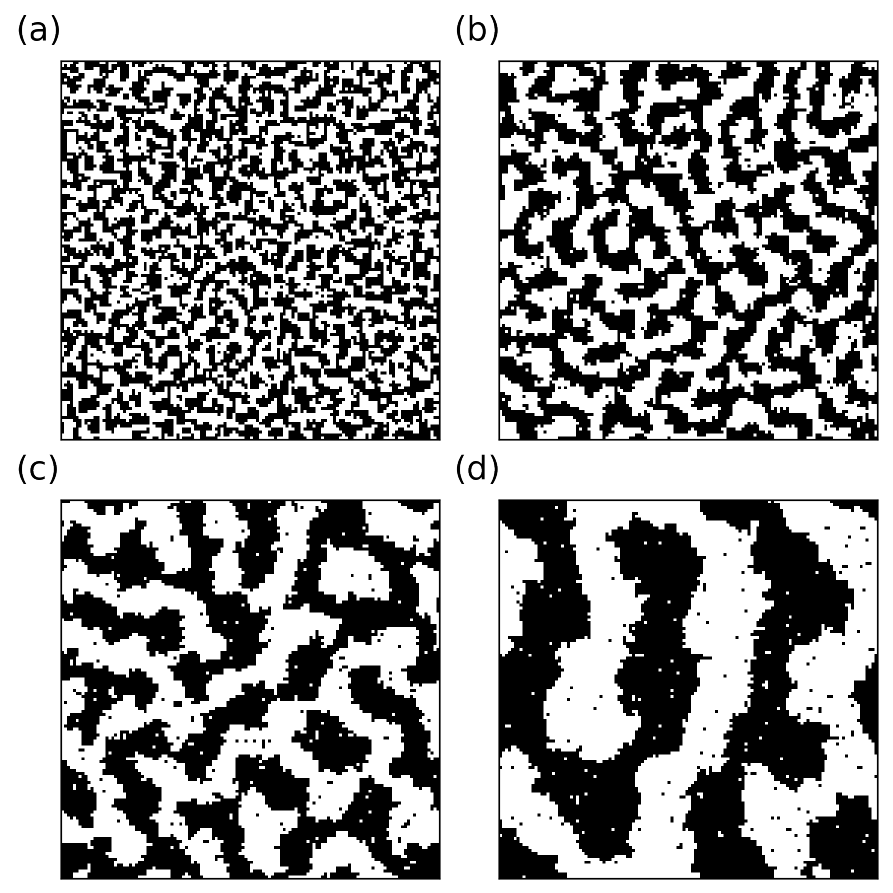}
    \caption{Evolution of spins in the dipole-conserving case, with system size $L=128$ and temperature $T = 0.75 \, T_c$. The snapshots are taken at Monte Carlo times (a) $10^4$, (b) $10^6$, (c) $10^8$, and (d) $10^{10}$.}
    \label{fig:growth}
\end{figure}

It is more convenient, however, to work with the Fourier transform of the correlator, the structure factor:
\begin{equation}
    \label{eq:structure}
    S(\bm{k}, t) = \langle \tilde{\sigma}_{\bm{k}}(t)  \tilde{\sigma}_{\bm{-k}}(t) \rangle,
\end{equation}
where $\tilde{\sigma}_{\bm{k}}(t) = \sum_{\bm{r}} e^{i \bm{k} \cdot \bm{r}} \sigma_{\bm{r}}(t)$ is the Fourier transform of the spin.  In our simulations, the structure factor is averaged over 100 independent trials, as well as over all lattice site pairs separated by $\bm{r}$.
The structure factor also obeys the scaling hypothesis, with late-time behavior
\begin{equation}\label{eq:structure_scaling}
S(\bm{k}, t) \approx R(t)^2 g(k R(t)),
\end{equation} 
where $g(x)$ is a scaling function.
Moreover, $S(\bm{k}, t)$ can be measured in scattering experiments~\cite{Bray1994review} and, unlike the correlation function, is always positive. Because of this latter property, we extract the typical domain size via the first radial moment of the structure factor~\cite{Amar1988}:
\begin{equation}\label{eq:structure_moment}
\expval{k(t)} = \frac{\sum_{\bm{k}} k \, S(\bm{k}, t)}{\sum_{\bm{k}} S(\bm{k}, t)}.
\end{equation}
Here the sums are taken over lattice momenta $|\bm{k}| \leq \pi$ in the Brillouin zone, which provides a natural cutoff. From Eq.~\eqref{eq:structure_moment}, the domain size is defined as 
\begin{equation}\label{eq:domain_moment}
    R(t) = \frac{2\pi}{\expval{k(t)}}.
\end{equation}
Aside from an overall prefactor that we ignore, Eq.~\eqref{eq:domain_moment} captures the characteristic size of domains at time $t$. 

At finite times, the power-law domain growth proceeds as $R(t) \sim t^{\zeta(t)}$, where we define the instantaneous exponent 
\begin{equation}
    \label{eq:coarsening_exp}
    \zeta(t) = \frac{d (\log R(t))}{d(\log t)}.
\end{equation}
As $R(t) \rightarrow \infty$, we expect $\zeta (t) \rightarrow 1/z$, where $z$ is the theoretical coarsening exponent. However, $\zeta (t)$ also reflects the early time corrections to scaling discussed in Sec.~\ref{sec:early_time}. 

\subsection{Numerical results}
\label{sec:num_results}
We now show that our Monte Carlo simulations agree with the analytical prediction $z = 2m+3$, and that the early-time corrections are captured by the exponent $z=2m+3+2^m$, as described in Sec.~\ref{sec:early_time}.

\begin{figure}[t!]
    \centering
    \hspace{-0.4cm}
    \includegraphics[width=\linewidth]{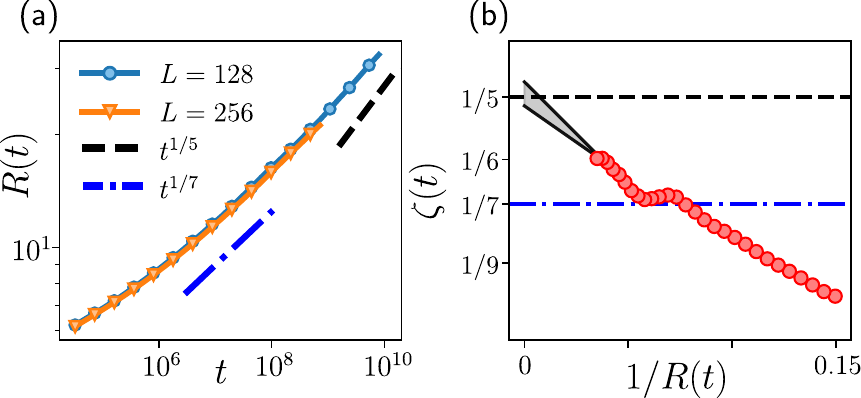}
    \caption{(a) Domain growth for dipole-conserving dynamics, as captured by $R(t)$. (b) The running logarithmic derivative of $R(t)$ clearly signals a region of early-time scaling $R(t) \sim t^{1/7}$ (blue dashed-dotted line), before pointing towards the asymptotic regime $R(t) \sim t^{1/5}$ (black dashed line). The shaded grey region between the solid black lines indicates different linear fits to the last 4-12 data points.
    }
    \label{fig:dip}
\end{figure}

First, we simulate ordinary Kawasaki dynamics in absence of dipolar or multipolar constraints. The resulting domain sizes and growth exponents are shown in Fig.~\ref{fig:Kawasaki}.
Only at long times is the asymptotic scaling $R(t) \sim t^{1/3}$ found, while at shorter times there are strong corrections to the growth law ~\cite{Huse1986}. This benchmark suggests that more constrained dynamics necessitate even longer times to access the asymptotic scaling regime.

We then move on to the results for dipole-conserving dynamics. In Fig.~\ref{fig:growth}, snapshots of the spin configurations at different times are shown. The late times required to observe large domains, as compared to Kawasaki dynamics, highlight the slower dynamics. Moreover, configurations always remain ``balanced'', as the center of mass of the magnetization is conserved in both spatial directions. 

In Fig.~\ref{fig:dip}, the growth of $R(t)$ is shown more quantitatively. While the early-time coarsening with $z=7$ is convincingly captured by a plateau in $\zeta(t)$, 
the true asymptotic scaling with $z=5$ is not obtained even for the longest times we could access. After $10^{10}$ Monte Carlo steps, the apparent coarsening exponent is around $z=6$, possibly reflecting the intermediate time regime discussed in Sec.~\ref{sec:early_time}. In Fig.~\ref{fig:dip}(b), a series of different linear extrapolations to $R(t) \rightarrow \infty$ are consistent with the late-time $z=5$ growth. 

\begin{figure}[t!]
    \centering
    \hspace{-0.4cm}
    \includegraphics[width=\linewidth]{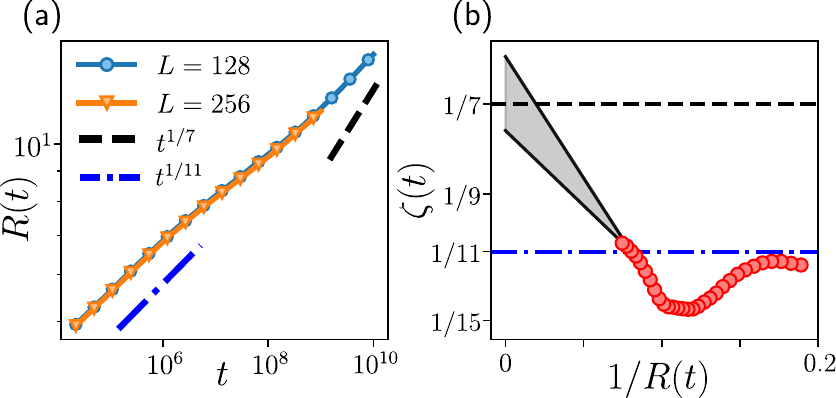}
    \caption{(a) Domain growth for quadrupole-conserving dynamics, as captured by $R(t)$. Even at the longest simulated times, the growth remains slower than the predicted $R(t) \sim t^{1/7}$. (b) The running logarithmic derivative of $R(t)$ indicates that finite-time corrections are likely scaling as $\sim t^{1/11}$, and that the asymptotic scaling is compatible with $z=7$. The shaded gray region between the solid black lines indicates different linear fits to the last 4-12 data points.
    }
    \label{fig:quad}
\end{figure}

Next, in Fig.~\ref{fig:quad} we show results for the quadrupole-conserving dynamics. Here, a clear verification of the asymptotic scaling $z=7$ was not possible with our simulations, as more than seven decades of additional Monte Carlo time would in principle be needed \emph{after the early-time corrections vanish}. While this daunting task remains outside reach, linear extrapolations from the largest accessible values of $R(t)$ yields a value of $z$ that is compatible with the analytical prediction $z=7$. Furthermore, early-time corrections appear to be well-described by a regime with $z=11$.

In App.~\ref{app:collapse}, we use our extracted domain sizes to self-consistently verify the dynamical scaling hypothesis. We collapse the correlation functions for different times onto the same scaling function, $C(r, t) = f(r/R(t))$, showing that a single length scale controls the growth of domains for several decades of Monte Carlo steps. Finally, we investigate temperature dependence in App.~\ref{app:temp}, finding that domain growth with a dipole constraint is similar for a range of temperatures. However, we leave questions of low-temperature jamming~\cite{Morningstar2020, Pozderac_2023, Cornell1991, Ben-Naim1998} to future work.

\section{Can multipole-conserving systems grow extensive domains?} \label{sec:proof}

Throughout this work, we have implicitly assumed that arbitrarily large domains can grow in multipole-conserving systems. Then, using that the motion of spins in such models is subdiffusive, we have determined the growth laws analytically in Sec.~\ref{sec:analytics} and confirmed them numerically in Sec.~\ref{sec:num}. The first assumption, however, is not trivial: multipole-conserving dynamics are known to cause Hilbert space fragmentation, so not all spins configurations are dynamically reachable one from another. Here, we prove that in the case of \emph{weak} fragmentation, the largest dynamical sector contains arbitrarily large domains, thus allowing for an asymptotic growth of $R(t)$ starting from an infinite-temperature state.

To build some intuition, let us first consider the one-dimensional case. In systems governed by Kawasaki dynamics, only spins located at a domain wall can be exchanged. Once a spin is inside an opposite domain, it can diffuse within it. Through this process, entire domains translate and eventually merge to form larger clusters. By performing local spin-exchange moves, the system can explore the entire configuration space at fixed total magnetization, which includes states with extensively large domains. For instance, at $M=0$, the largest domain in a system of size $L$ has size $L/2$.

Now consider the case where dipole conservation is enforced. In this scenario, isolated charges become immobile, manifesting the fractonic nature of the system. For instance, a down spin embedded in a sea of up spins,
\begin{equation}
    \uparrow^{m_1} \downarrow \uparrow^{m_2},
\end{equation}
cannot propagate through the surrounding domain, as it would in Kawasaki dynamics, because doing so would alter the system's dipole moment\footnote{We use the notation $\sigma^{m}$ to denote a configuration of $m$ adjoining spins $\sigma$.}. 

The simplest allowed updates in dipole-conserving dynamics are of the form
\begin{equation}
\begin{aligned}
    \uparrow \downarrow \downarrow \uparrow \quad &\longleftrightarrow \quad \downarrow \uparrow \uparrow \downarrow, \\
    \u \d \sigma \d \u \quad &\longleftrightarrow \quad \d \u \sigma \u \d,
\end{aligned}
\end{equation}
where the two updates have respective ranges $r=4$ and $r=5$, and the longer-range term is necessary to avoid strong fragmentation~\cite{Sala2020, Moudgalya2020}.
We can interpret the local dipole configurations as effective dipole ``particles'': $\uparrow \downarrow$ represents a left-pointing dipole $(\Leftarrow)$, while $\downarrow \uparrow$ represents a right-pointing dipole $(\Rightarrow)$. The dynamics then consists of exchanging these local dipoles while conserving the total dipole moment.

Crucially, there are an exponentially large number of configurations that are completely frozen, i.e., dynamically inaccessible, due to the conservation laws. These configurations are present for both weak and strong fragmentation, 
and in the quantum case correspond to zero modes and quantum scars~\cite{Sala2022, Chandran2023}.
Consider, for example, a series of alternating domains whose sizes $m_i$ are larger than the maximum spin-exchange range $r$,
\begin{equation}
    \uparrow^{m_1} \downarrow^{m_2} \uparrow^{m_3} \downarrow^{m_4} \cdots.
\end{equation}
These configurations are frozen because spins inside the domains cannot flip: they already locally extremize the magnetization by being fully aligned (all up or all down). 
At the domain interfaces, the configuration also extremizes the local dipole moment, preventing any further moves without violating the conservation laws. Clearly, the number of such frozen configurations grows exponentially with the system size; in fact, the system contains at least $2^{L/r}$ frozen states.

Energetically, these configurations tend to have low energy in systems with ferromagnetic Ising-type interactions, as they are composed of large aligned domains. Thus, there is a clear competition between thermal fluctuations at low temperatures, which favor the growth of domains, and dipole (or more generally, multipole) conservation, which tends to freeze states with domains that are too large.
This observation naturally raises the question of whether the largest connected set of configurations can still support large domains. In particular, we ask whether such domains can (1) be larger than the update range $r$, and (2) scale extensively with the system size $L$.
Below, we answer both of these questions in the affirmative. 

To prove that domains can be arbitrarily large, we consider the one-dimensional case. The result then automatically extends to higher dimensions, as domains are allowed to grow arbitrarily large in each direction independently. 
Let us first compute the number of states in which the largest domain has size at most $S$. This problem is equivalent to counting the number of ways to partition an integer $L$ into integers $\{n_i\}$ such that $1 \leq n_i \leq S$\footnote{Since the order of our $\{n_i\}$ matters, we are technically counting compositions instead of partitions. Moreover, the number of spin configurations with domains of length at most $S$ is actually twice the number of restricted compositions, since one can start with either an up or a down spin.}, i.e.,
\begin{equation}
    \sum_{i=1}^{\infty} n_i = L, \quad \text{with } 1 \leq n_i \leq S.
\end{equation} 
The corresponding generating function is
\begin{equation}\label{eq:generating}
    G(z) = \sum_{L=0}^{\infty} \big(z + \dots + z^S \big)^L = \frac{1-z}{1-2z+z^{S+1}}.
\end{equation}
Expanding the generating function as 
\begin{equation}
    G(z) = \sum_{L=0}^{\infty} a_L z^L,
\end{equation}
we identify the coefficient $a_L$ as the number of states with domains of size at most $S$ in a system of size $L$.

The asymptotic behavior of $a_L$ as $L\rightarrow \infty$ is controlled by the singularities of its generating function in the complex plane~\cite{flajolet}.
The function $G(z)$ is analytic in a neighborhood of $z=0$ until it encounters its first pole at $z=\rho$, where it behaves as 
\begin{equation}
    G(z) \sim  \left(1-\frac{z}{\rho}\right)^{\eta}.
\end{equation}
According to the Flajolet–Odlyzko Theorem, the generating function coefficients asymptotically scale as 
\begin{equation}
    a_L \sim \rho^{-L} L^{-\eta -1}.
\end{equation}
In the limit $S \rightarrow \infty$, the first pole $\rho$ approaches $1/2$ from above. To find the corrections for large but finite $S$, we set $\rho = \frac{1}{2} + \delta$ and expand in $\delta$.
Assuming $\delta \ll 1$, the solution behaves as
\begin{equation}
    \delta = \frac{1}{2} \left(\frac{1}{2} + \delta\right)^S \approx 2^{-S-1}.
\end{equation}
As a result,
\begin{equation}
    a_L \sim  2^{\gamma L}, \quad     \gamma = S+1-\log_2(2^S + 1).
\end{equation}
where crucially $\gamma<1$ for any fixed $S$. 

The number of states with domains of length at most $S$ must now be compared with the number of states in the largest connected component. The size of the latter scales as $|\mathcal{C}_\text{largest}| \sim 2^L / L^{\alpha}$ for weak fragmentation, where the denominator is a polynomial correction that depends on the conserved charges. 
At large enough system sizes, ${2^L}/{L^{\alpha}} > 2^{\gamma L}$, and the connected sector necessarily contains states with domains larger than any fixed $S$. This ensures that domains can grow arbitrarily large during dynamics. 

Although the above proof is elegant and simple, its main limitation is that it does not account for the detailed structure of the dynamics. It merely states that the number of states with domains smaller than a fixed $S$ is too small, so the largest dynamical sector must contain states with larger domains. As a result, the maximum domain size grows without bound as $L$ increases, but it could scale more slowly than $\mathcal{O}(L)$.

We now prove a stronger result: for any $L$, there exist configurations in the largest connected component with a domain larger than $\nu L$, where $\nu$ is a fixed constant. Consequently, as long as the exchange range $r$ is sufficiently large, configurations with extensive domains walls are accessible to the dynamics. 
The starting point is the definition of weak fragmentation, which implies
\begin{equation}
    |\mathcal{C}_{P = 0}| - |\mathcal{C}_\text{largest}| \sim 2^{\mu L}, \quad \mu < 1,
\end{equation} where $\mathcal{C}_{P=0}\supset \mathcal{C}_\text{largest}$ is shorthand to denote the symmetry sector with all conserved multipole moments set to zero. In other words, the size of the largest connected component differs from the full configuration space by an exponentially small fraction. 

Suppose that the $\mathcal{C}_\text{largest}$ does not contain any configurations with a domain larger than $\nu L$. 
This would imply that all configurations with domains larger than $\nu L$ must belong to the complement of $\mathcal{C}_\text{largest}$ in $\mathcal{C}_{P=0}$.  
Now consider the number of configurations with $P^{(0, 1, \ldots, m)} = {0}$ and at least one domain of size larger than $\nu L$. If this number exceeds $\sim 2^{\mu L}$, we would reach a contradiction. By the pigeonhole principle, some of these large-domain configurations must therefore belong to $\mathcal{C}_\text{largest}$.

It is straightforward to construct an exponentially large set of such configurations to obtain the contradiction. For simplicity, we consider the dipole case ($m=1$); the generalization to higher multipoles is immediate. Take a completely random configurations of spins on a chain of length $L/4$,
\begin{equation}
\Sigma = \sigma_1 \sigma_2 \ldots \sigma_{L/4}.
\end{equation} 
There are $2^{L/4}$ such configurations, and we can construct a state with $P^{(0,1)} = 0$ by replicating one of these configurations four times. Specifically, the configuration $(+\Sigma)(-\Sigma) (-\Sigma) (+\Sigma)$ has length $L$ and no magnetization or dipole moment. 

Within this symmetry sector, we would like to lower bound the number of states with a domain of size at least $\nu L$. An obvious way to build such a state is by starting with a smaller system $\Sigma$ that already has a domain of length $\nu L$, and then replicating it. Since there are $L/4 - \nu L$ remaining free spins in $\Sigma$, the number of such configurations is lower bounded by $2^{L/4 - \nu L}$. 
The condition to obtain the contradiction is then
\begin{equation}
\frac{1}{4} - \nu > \mu.
\end{equation}
If this inequality is satisfied, we can conclude that configurations 
with a domain larger than $\nu L$ outnumber the configurations in $\mathcal{C}_{P=0} \setminus \mathcal{C}_\text{largest}$. As a result, these extensive-domain states must bleed into $\mathcal{C}_\text{largest}$. 

The value of $\mu$ depends on the exchange range $r$: the larger $r$ is, the less fragmented the system becomes. In the limit of large $r$, the largest connected component coincides with the entire symmetry sector, and $\mu\rightarrow 0$. Therefore, by choosing a sufficiently large exchange range $r$, one can satisfy the condition $1/4 - \nu > \mu(r)$. As a result, exponentially many accessible configurations in $\mathcal{C}_\text{largest}$ contain domains of size larger than $\nu L$.

\section{Conclusion}\label{sec:conclusion}
In this work, we studied coarsening in spin systems with fractonic mobility constraints. Following a quench into the ordered phase of the Ising model, we tracked how the growth of domains is modified when higher moments of the order parameter are conserved, such as the dipole moment.
Though we focused on spin systems, our results hold for conserved multipoles of any scalar order parameter.
It is well known that the linear size of domains grows as $R(t) \sim t^{1/2}$ when the order parameter is not conserved (Glauber dynamics), and $R(t) \sim t^{1/3}$ when the order parameter is conserved (Kawasaki dynamics). By generalizing this theory to the multipole case, we showed that if the $m$-th moment of the order parameter is conserved, then domains grow as $R(t) \sim t^{1/(2m+3)}$.

The dynamics of systems with $m$-pole conservation are highly constrained when $m \geq 1$, raising the question of whether domains can grow indefinitely with system size.
Even in the simplest case of dipole-conservation, regions with large domains are nearly frozen, as spins cannot exchange across domain walls without altering the total dipole moment.
This behavior should be contrasted with Kawasaki dynamics, in which spins at domain walls are mobile. Nevertheless, for the time scales that we explored, an unbounded domain growth was observed.

Furthermore, we provided a simple combinatorial argument that, under mild assumptions, shows that states with extensively large domains are dynamically accessible.
While domains scale with system size at asymptotically long times, growth may be problematic if the thermodynamic limit is taken first, since the relevant time scales might diverge. 
This allows for ergodicity-breaking and aging phenomena in such systems at finite temperature, a natural direction for future investigation. Indeed, previous work on coarsening with kinetic constraints uncovered similar glassy phenomena~\cite{Sollich1999, Sollich2003}, which may also extend to classical fracton spin liquids~\cite{Placke2024Jul}.

This work opens other natural directions for further research, including the study of coarsening in systems with modulated symmetries, the effect of long-range updates on the dynamics, and the role of defects that weakly break the conservation laws. 
Among the most interesting directions is the study of coarsening in multipole-conserving \emph{quantum} systems~\cite{Turkowski2006,Aron2009, Chandran2013, Maraga2015, Samajdar2024, Balducci2025} 
as recent experiments have shown the capability of quantum simulators to access the coarsening regime in isolated systems undergoing unitary dynamics~\cite{Andersen2025Thermalization,Manovitz2025Quantum}. By quenching or ramping the system from a disordered to an ordered phase, one can seed the phase ordering kinetics in quantum simulators and measure the dynamical exponent $z$ via imaging. 
Moreover, dipole conservation can be realized as an emergent symmetry in cold-atom systems subject to a tilted external potential~\cite{Guardado-Sanchez2020, Scherg2021, Kohlert2023}. 
Whether the interplay of quantum fluctuations and constraints would lead to localization is a difficult and deep research question that deserves to be addressed.
 
\textit{Acknowledgements}.---We thank Soumya Bera, Taylor L. Hughes, and Pablo Sala for several illuminating discussions. We thank David Huse for a critical reading of the manuscript 
and for providing important feedback.
J.G. acknowledges support from the US Office of Naval Research MURI grant N00014-20-1-2325. G.D.T. acknowledges support from the EPiQS Program of the Gordon and Betty Moore Foundation and the hospitality of MPIPKS Dresden, where part of the work was done. 

\appendix
\section{Chemical potential near a domain wall}
\label{app:gibbsthomson}
In this Appendix, we derive the Gibbs-Thomson relation (Eq.~\eqref{eq:gibbsthomson} in the main text) for the chemical potential near a domain wall. For convenience, we reproduce the Landau-Ginzburg free energy below:
\begin{equation}\label{eq:app_lg}
        F[\phi] = \int d^d x \left[\frac{1}{2}(\nabla \phi)^2 + V(\phi)\right].
\end{equation}
Working near a domain wall, we denote the direction perpendicular to the interface by the coordinate $r$. The profile of the order parameter near the domain wall can be taken to be $\phi(r) \sim \tanh(r)$, with $\phi(r=\pm \infty) = \pm 1$ and $\phi(0) = 0$.
\begin{figure}[t!]
    \centering
    \includegraphics[width=\linewidth]{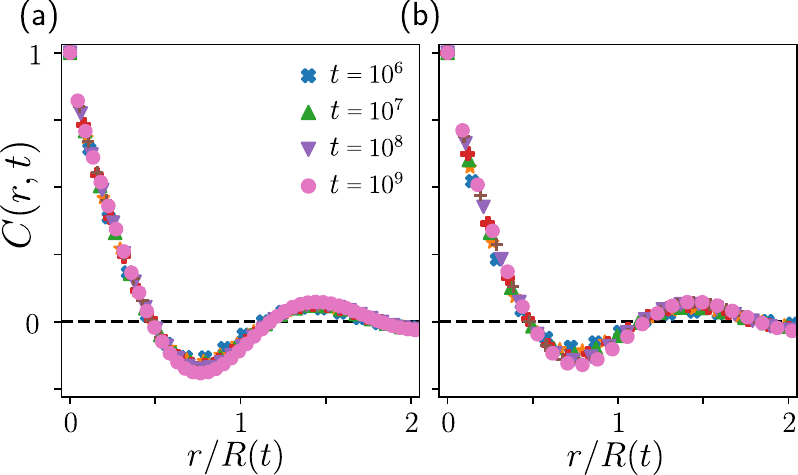}
    \caption{Correlation function $C(r, t)$ for (a) dipole- and (b) quadrupole-conserving dynamics on a lattice with $L=256$. The data are plotted for versus rescaled distances $r/R(t)$, where $R(t)$ is extracted numerically from Eq.~\eqref{eq:domain_moment}. Seven different times between $10^6$ and $10^9$ are shown and collapse reasonably well onto the same curve.
    }
    \label{fig:collapse}
\end{figure}

From the Landau-Ginzburg free energy, we extract the chemical potential $\mu = \delta F/\delta \phi$, yielding
\begin{equation}\label{eq:app_mu}
    \mu = - \partial_r^2 \phi - (\partial_r \phi)(\nabla \cdot \hat{\bm{r}}) + V''(\phi),
\end{equation}
where $\hat{\bm{r}}$ is the unit normal to the domain wall.
In this expression, we can recognize $\kappa = \nabla \cdot \hat{\bm{r}}$ as the extrinsic curvature of the interface. 

Next, we multiply both sides of Eq.~\eqref{eq:app_mu} by $\partial_r \phi$ and integrate across the domain wall. Because $\partial_r \phi$ is only nonzero in the immediate vicinity of the interface, we can safely extend our integration bounds to infinity. On the left-hand side, we are left with
\begin{equation}
\int_{-\infty}^{\infty} dr \mu (\partial_r \phi) \approx \mu_\text{avg} \Delta \phi
\end{equation}
where $\Delta \phi$ is the change in order-parameter across the domain wall, and $\mu_\text{avg}$ is the average chemical potential at the interface.
On the right hand side, we have
\begin{figure}[t!]
    \centering
    \includegraphics[width=0.8\linewidth]{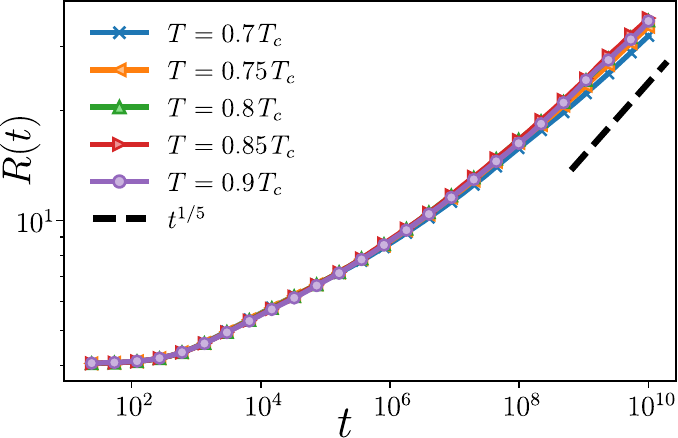}
    \caption{Growth of domains for dipole-conserving dynamics at different temperatures. For the range of temperatures shown ($T = 0.7\, T_c$ to $T = 0.9\, T_c$), the coarsening is approximately temperature-independent.
    }
    \label{fig:temp}
\end{figure}
\begin{equation}\label{eq:app_rhs}
\begin{aligned}
&\int_{-\infty}^{\infty}dr \left[ -(\partial_r \phi) \partial_r^2 \phi - (\partial_r \phi)^2 (\nabla \cdot \hat{\bm{r}}) + (\partial_r \phi) V'(\phi) \right]\\
&= -\int (\partial_r \phi)^2 (\nabla \cdot \hat{\bm{r}})  \approx -\gamma \kappa
\end{aligned}
\end{equation}
The first term in Eq.~\eqref{eq:app_rhs} vanishes because it is an integral of an odd quantity. The last term becomes $\int dr \partial_r V = \Delta V$, which is zero because the double-well potential takes the same value inside both domains. 
The middle term of Eq.~\eqref{eq:app_rhs} is therefore the only one that does not vanish. 

Since we are integrating in the vicinity of the domain wall, we can treat the curvature as approximately constant. If $\phi$ only varies across the domain wall, the remaining integral, $\int dr (\partial_r \phi)^2$, is then the surface tension, $\gamma = \int d^d x |\nabla \phi|^2$. Putting these all together, we find the Gibbs-Thomson law:
\begin{equation}
    \mu_\text{avg} = -\frac{\gamma \kappa}{\Delta \phi}
\end{equation}

\section{Scaling collapse of correlation functions}\label{app:collapse}
In this Appendix, we show additional numerical data to support the dynamical scaling hypothesis that there is a single length scale in coarsening dynamics. If this hypothesis holds, then the correlation function should have an approximate scaling form, Eq.~\eqref{eq:scaling}.
Here, we provide evidence for this scaling in case of dipole and quadrupole conserving dynamics. In Fig.~\ref{fig:collapse}, we plot the correlation function $C(r,t)$ at several different times, and show that when the distance is rescaled by our numerically computed $R(t)$, the data collapse reasonably well on the same curve.

The correlation function is computed by averaging over spacings  $\bm{r}$ parallel to the horizontal and vertical axes of the lattice, as well as over 100 different trials. We plot data at 7 times ranging from $10^6$ to $10^9$ Monte Carlo steps, taken from  simulations of a $256 \times 256$ lattice.

\section{Temperature dependence of coarsening}\label{app:temp}
In this Appendix, we present numerical results for coarsening at temperatures different from $T= 0.75 \, T_c$, which was used in the main text. We focus on the case of dipole-conservation for simplicity. To maintain a separation of scales between the domain size and the equilibrium correlation length, we consider a range of temperatures from $T = 0.7\, T_c$ to $T = 0.9\, T_c$. The average size of domains, $R(t)$, is plotted for these temperatures in Fig.~\ref{fig:temp}. Despite mild variations, domain sizes are nearly identical for different values of $T$, reflecting the irrelevance of temperature for sub-critical coarsening~\cite{Bray1994review}.

We have avoided the low temperature regime because of the severity of early-time corrections, leaving the interesting question of a temperature-induced fragmentation or freezing~\cite{Morningstar2020, Pozderac_2023} transition to future work. However, even ordinary Kawasaki dynamics features jamming at $T=0$~\cite{Cornell1991, Ben-Naim1998}, so it is natural to expect similar behavior in a system with ulterior kinetic constraints.

\bibliographystyle{apsrev4-2}
\bibliography{main.bib}

\begin{thebibliography}{67}%
\makeatletter
\providecommand \@ifxundefined [1]{%
 \@ifx{#1\undefined}
}%
\providecommand \@ifnum [1]{%
 \ifnum #1\expandafter \@firstoftwo
 \else \expandafter \@secondoftwo
 \fi
}%
\providecommand \@ifx [1]{%
 \ifx #1\expandafter \@firstoftwo
 \else \expandafter \@secondoftwo
 \fi
}%
\providecommand \natexlab [1]{#1}%
\providecommand \enquote  [1]{``#1''}%
\providecommand \bibnamefont  [1]{#1}%
\providecommand \bibfnamefont [1]{#1}%
\providecommand \citenamefont [1]{#1}%
\providecommand \href@noop [0]{\@secondoftwo}%
\providecommand \href [0]{\begingroup \@sanitize@url \@href}%
\providecommand \@href[1]{\@@startlink{#1}\@@href}%
\providecommand \@@href[1]{\endgroup#1\@@endlink}%
\providecommand \@sanitize@url [0]{\catcode `\\12\catcode `\$12\catcode `\&12\catcode `\#12\catcode `\^12\catcode `\_12\catcode `\%12\relax}%
\providecommand \@@startlink[1]{}%
\providecommand \@@endlink[0]{}%
\providecommand \url  [0]{\begingroup\@sanitize@url \@url }%
\providecommand \@url [1]{\endgroup\@href {#1}{\urlprefix }}%
\providecommand \urlprefix  [0]{URL }%
\providecommand \Eprint [0]{\href }%
\providecommand \doibase [0]{https://doi.org/}%
\providecommand \selectlanguage [0]{\@gobble}%
\providecommand \bibinfo  [0]{\@secondoftwo}%
\providecommand \bibfield  [0]{\@secondoftwo}%
\providecommand \translation [1]{[#1]}%
\providecommand \BibitemOpen [0]{}%
\providecommand \bibitemStop [0]{}%
\providecommand \bibitemNoStop [0]{.\EOS\space}%
\providecommand \EOS [0]{\spacefactor3000\relax}%
\providecommand \BibitemShut  [1]{\csname bibitem#1\endcsname}%
\let\auto@bib@innerbib\@empty
\bibitem [{\citenamefont {Bray}(1994)}]{Bray1994review}%
  \BibitemOpen
  \bibfield  {author} {\bibinfo {author} {\bibfnamefont {A.~J.}\ \bibnamefont {Bray}},\ }\href {https://www.tandfonline.com/doi/abs/10.1080/00018739400101505} {\bibfield  {journal} {\bibinfo  {journal} {Adv. Phys.}\ } (\bibinfo {year} {1994})}\BibitemShut {NoStop}%
\bibitem [{\citenamefont {Krapivsky}\ \emph {et~al.}(2010)\citenamefont {Krapivsky}, \citenamefont {Redner},\ and\ \citenamefont {Ben\-Naim}}]{Krapivsky}%
  \BibitemOpen
  \bibfield  {author} {\bibinfo {author} {\bibfnamefont {P.~L.}\ \bibnamefont {Krapivsky}}, \bibinfo {author} {\bibfnamefont {S.}~\bibnamefont {Redner}},\ and\ \bibinfo {author} {\bibfnamefont {E.}~\bibnamefont {Ben\-Naim}},\ }\href@noop {} {\emph {\bibinfo {title} {{A Kinetic View of Statistical Physics}}}}\ (\bibinfo  {publisher} {Cambridge University Press},\ \bibinfo {year} {2010})\BibitemShut {NoStop}%
\bibitem [{\citenamefont {Berges}\ and\ \citenamefont {Hoffmeister}(2009)}]{Berges2009}%
  \BibitemOpen
  \bibfield  {author} {\bibinfo {author} {\bibfnamefont {J.}~\bibnamefont {Berges}}\ and\ \bibinfo {author} {\bibfnamefont {G.}~\bibnamefont {Hoffmeister}},\ }\href {https://doi.org/10.1016/j.nuclphysb.2008.12.017} {\bibfield  {journal} {\bibinfo  {journal} {Nuclear Physics B}\ }\textbf {\bibinfo {volume} {813}},\ \bibinfo {pages} {383} (\bibinfo {year} {2009})}\BibitemShut {NoStop}%
\bibitem [{\citenamefont {Glauber}(1963)}]{Glauber1963}%
  \BibitemOpen
  \bibfield  {author} {\bibinfo {author} {\bibfnamefont {R.~J.}\ \bibnamefont {Glauber}},\ }\href {https://doi.org/10.1063/1.1703954} {\bibfield  {journal} {\bibinfo  {journal} {J. Math. Phys.}\ }\textbf {\bibinfo {volume} {4}},\ \bibinfo {pages} {294} (\bibinfo {year} {1963})}\BibitemShut {NoStop}%
\bibitem [{\citenamefont {Kawasaki}(1966)}]{Kawasaki1966}%
  \BibitemOpen
  \bibfield  {author} {\bibinfo {author} {\bibfnamefont {K.}~\bibnamefont {Kawasaki}},\ }\href {https://doi.org/10.1103/PhysRev.145.224} {\bibfield  {journal} {\bibinfo  {journal} {Phys. Rev.}\ }\textbf {\bibinfo {volume} {145}},\ \bibinfo {pages} {224} (\bibinfo {year} {1966})}\BibitemShut {NoStop}%
\bibitem [{\citenamefont {Nandkishore}\ and\ \citenamefont {Hermele}(2019)}]{Nandkishore_2019}%
  \BibitemOpen
  \bibfield  {author} {\bibinfo {author} {\bibfnamefont {R.~M.}\ \bibnamefont {Nandkishore}}\ and\ \bibinfo {author} {\bibfnamefont {M.}~\bibnamefont {Hermele}},\ }\href {https://doi.org/https://doi.org/10.1146/annurev-conmatphys-031218-013604} {\bibfield  {journal} {\bibinfo  {journal} {Annual Review of Condensed Matter Physics}\ }\textbf {\bibinfo {volume} {10}},\ \bibinfo {pages} {295} (\bibinfo {year} {2019})}\BibitemShut {NoStop}%
\bibitem [{\citenamefont {Vijay}\ \emph {et~al.}(2015)\citenamefont {Vijay}, \citenamefont {Haah},\ and\ \citenamefont {Fu}}]{Vijay_2015}%
  \BibitemOpen
  \bibfield  {author} {\bibinfo {author} {\bibfnamefont {S.}~\bibnamefont {Vijay}}, \bibinfo {author} {\bibfnamefont {J.}~\bibnamefont {Haah}},\ and\ \bibinfo {author} {\bibfnamefont {L.}~\bibnamefont {Fu}},\ }\href {https://doi.org/10.1103/PhysRevB.92.235136} {\bibfield  {journal} {\bibinfo  {journal} {Phys. Rev. B}\ }\textbf {\bibinfo {volume} {92}},\ \bibinfo {pages} {235136} (\bibinfo {year} {2015})}\BibitemShut {NoStop}%
\bibitem [{\citenamefont {Vijay}\ \emph {et~al.}(2016)\citenamefont {Vijay}, \citenamefont {Haah},\ and\ \citenamefont {Fu}}]{Vijay_2016}%
  \BibitemOpen
  \bibfield  {author} {\bibinfo {author} {\bibfnamefont {S.}~\bibnamefont {Vijay}}, \bibinfo {author} {\bibfnamefont {J.}~\bibnamefont {Haah}},\ and\ \bibinfo {author} {\bibfnamefont {L.}~\bibnamefont {Fu}},\ }\href {https://doi.org/10.1103/PhysRevB.94.235157} {\bibfield  {journal} {\bibinfo  {journal} {Phys. Rev. B}\ }\textbf {\bibinfo {volume} {94}},\ \bibinfo {pages} {235157} (\bibinfo {year} {2016})}\BibitemShut {NoStop}%
\bibitem [{\citenamefont {Pretko}(2017)}]{Pretko2017}%
  \BibitemOpen
  \bibfield  {author} {\bibinfo {author} {\bibfnamefont {M.}~\bibnamefont {Pretko}},\ }\href {https://doi.org/10.1103/PhysRevB.95.115139} {\bibfield  {journal} {\bibinfo  {journal} {Phys. Rev. B}\ }\textbf {\bibinfo {volume} {95}},\ \bibinfo {pages} {115139} (\bibinfo {year} {2017})}\BibitemShut {NoStop}%
\bibitem [{\citenamefont {Pretko}\ and\ \citenamefont {Radzihovsky}(2018)}]{Pretko_2018}%
  \BibitemOpen
  \bibfield  {author} {\bibinfo {author} {\bibfnamefont {M.}~\bibnamefont {Pretko}}\ and\ \bibinfo {author} {\bibfnamefont {L.}~\bibnamefont {Radzihovsky}},\ }\href {https://doi.org/10.1103/PhysRevLett.120.195301} {\bibfield  {journal} {\bibinfo  {journal} {Phys. Rev. Lett.}\ }\textbf {\bibinfo {volume} {120}},\ \bibinfo {pages} {195301} (\bibinfo {year} {2018})}\BibitemShut {NoStop}%
\bibitem [{\citenamefont {Sala}\ \emph {et~al.}(2020)\citenamefont {Sala}, \citenamefont {Rakovszky}, \citenamefont {Verresen}, \citenamefont {Knap},\ and\ \citenamefont {Pollmann}}]{Sala2020}%
  \BibitemOpen
  \bibfield  {author} {\bibinfo {author} {\bibfnamefont {P.}~\bibnamefont {Sala}}, \bibinfo {author} {\bibfnamefont {T.}~\bibnamefont {Rakovszky}}, \bibinfo {author} {\bibfnamefont {R.}~\bibnamefont {Verresen}}, \bibinfo {author} {\bibfnamefont {M.}~\bibnamefont {Knap}},\ and\ \bibinfo {author} {\bibfnamefont {F.}~\bibnamefont {Pollmann}},\ }\href {https://doi.org/10.1103/PhysRevX.10.011047} {\bibfield  {journal} {\bibinfo  {journal} {Physical Review X}\ }\textbf {\bibinfo {volume} {10}},\ \bibinfo {pages} {011047} (\bibinfo {year} {2020})}\BibitemShut {NoStop}%
\bibitem [{\citenamefont {Khemani}\ \emph {et~al.}(2020)\citenamefont {Khemani}, \citenamefont {Hermele},\ and\ \citenamefont {Nandkishore}}]{Khemani2020}%
  \BibitemOpen
  \bibfield  {author} {\bibinfo {author} {\bibfnamefont {V.}~\bibnamefont {Khemani}}, \bibinfo {author} {\bibfnamefont {M.}~\bibnamefont {Hermele}},\ and\ \bibinfo {author} {\bibfnamefont {R.}~\bibnamefont {Nandkishore}},\ }\href {https://doi.org/10.1103/PhysRevB.101.174204} {\bibfield  {journal} {\bibinfo  {journal} {Phys. Rev. B}\ }\textbf {\bibinfo {volume} {101}},\ \bibinfo {pages} {174204} (\bibinfo {year} {2020})}\BibitemShut {NoStop}%
\bibitem [{\citenamefont {Gromov}\ \emph {et~al.}(2020)\citenamefont {Gromov}, \citenamefont {Lucas},\ and\ \citenamefont {Nandkishore}}]{Gromov2020}%
  \BibitemOpen
  \bibfield  {author} {\bibinfo {author} {\bibfnamefont {A.}~\bibnamefont {Gromov}}, \bibinfo {author} {\bibfnamefont {A.}~\bibnamefont {Lucas}},\ and\ \bibinfo {author} {\bibfnamefont {R.~M.}\ \bibnamefont {Nandkishore}},\ }\href {https://doi.org/10.1103/PhysRevResearch.2.033124} {\bibfield  {journal} {\bibinfo  {journal} {Physical Review Research}\ }\textbf {\bibinfo {volume} {2}},\ \bibinfo {pages} {033124} (\bibinfo {year} {2020})}\BibitemShut {NoStop}%
\bibitem [{\citenamefont {Feldmeier}\ \emph {et~al.}(2020)\citenamefont {Feldmeier}, \citenamefont {Sala}, \citenamefont {De~Tomasi}, \citenamefont {Pollmann},\ and\ \citenamefont {Knap}}]{Feldmeier2020}%
  \BibitemOpen
  \bibfield  {author} {\bibinfo {author} {\bibfnamefont {J.}~\bibnamefont {Feldmeier}}, \bibinfo {author} {\bibfnamefont {P.}~\bibnamefont {Sala}}, \bibinfo {author} {\bibfnamefont {G.}~\bibnamefont {De~Tomasi}}, \bibinfo {author} {\bibfnamefont {F.}~\bibnamefont {Pollmann}},\ and\ \bibinfo {author} {\bibfnamefont {M.}~\bibnamefont {Knap}},\ }\href {https://doi.org/10.1103/PhysRevLett.125.245303} {\bibfield  {journal} {\bibinfo  {journal} {Phys. Rev. Lett.}\ }\textbf {\bibinfo {volume} {125}},\ \bibinfo {pages} {245303} (\bibinfo {year} {2020})}\BibitemShut {NoStop}%
\bibitem [{\citenamefont {Ogunnaike}\ \emph {et~al.}(2023)\citenamefont {Ogunnaike}, \citenamefont {Feldmeier},\ and\ \citenamefont {Lee}}]{Ogunnaike_2023}%
  \BibitemOpen
  \bibfield  {author} {\bibinfo {author} {\bibfnamefont {O.}~\bibnamefont {Ogunnaike}}, \bibinfo {author} {\bibfnamefont {J.}~\bibnamefont {Feldmeier}},\ and\ \bibinfo {author} {\bibfnamefont {J.~Y.}\ \bibnamefont {Lee}},\ }\href {https://doi.org/10.1103/PhysRevLett.131.220403} {\bibfield  {journal} {\bibinfo  {journal} {Phys. Rev. Lett.}\ }\textbf {\bibinfo {volume} {131}},\ \bibinfo {pages} {220403} (\bibinfo {year} {2023})}\BibitemShut {NoStop}%
\bibitem [{\citenamefont {Morningstar}\ \emph {et~al.}(2023)\citenamefont {Morningstar}, \citenamefont {O'Dea},\ and\ \citenamefont {Richter}}]{Morningstar_2023}%
  \BibitemOpen
  \bibfield  {author} {\bibinfo {author} {\bibfnamefont {A.}~\bibnamefont {Morningstar}}, \bibinfo {author} {\bibfnamefont {N.}~\bibnamefont {O'Dea}},\ and\ \bibinfo {author} {\bibfnamefont {J.}~\bibnamefont {Richter}},\ }\href {https://doi.org/10.1103/PhysRevB.108.L020304} {\bibfield  {journal} {\bibinfo  {journal} {Phys. Rev. B}\ }\textbf {\bibinfo {volume} {108}},\ \bibinfo {pages} {L020304} (\bibinfo {year} {2023})}\BibitemShut {NoStop}%
\bibitem [{\citenamefont {Gliozzi}\ \emph {et~al.}(2023)\citenamefont {Gliozzi}, \citenamefont {May-Mann}, \citenamefont {Hughes},\ and\ \citenamefont {De~Tomasi}}]{Gliozzi2023}%
  \BibitemOpen
  \bibfield  {author} {\bibinfo {author} {\bibfnamefont {J.}~\bibnamefont {Gliozzi}}, \bibinfo {author} {\bibfnamefont {J.}~\bibnamefont {May-Mann}}, \bibinfo {author} {\bibfnamefont {T.~L.}\ \bibnamefont {Hughes}},\ and\ \bibinfo {author} {\bibfnamefont {G.}~\bibnamefont {De~Tomasi}},\ }\href {https://doi.org/10.1103/PhysRevB.108.195106} {\bibfield  {journal} {\bibinfo  {journal} {Phys. Rev. B}\ }\textbf {\bibinfo {volume} {108}},\ \bibinfo {pages} {195106} (\bibinfo {year} {2023})}\BibitemShut {NoStop}%
\bibitem [{\citenamefont {Lifshitz}\ and\ \citenamefont {Slyozov}(1961)}]{Lifshitz1961}%
  \BibitemOpen
  \bibfield  {author} {\bibinfo {author} {\bibfnamefont {I.}~\bibnamefont {Lifshitz}}\ and\ \bibinfo {author} {\bibfnamefont {V.}~\bibnamefont {Slyozov}},\ }\href {https://doi.org/10.1016/0022-3697(61)90054-3} {\bibfield  {journal} {\bibinfo  {journal} {J. Phys. Chem. Solids}\ }\textbf {\bibinfo {volume} {19}},\ \bibinfo {pages} {35} (\bibinfo {year} {1961})}\BibitemShut {NoStop}%
\bibitem [{\citenamefont {Wagner}(1961)}]{Wagner1961}%
  \BibitemOpen
  \bibfield  {author} {\bibinfo {author} {\bibfnamefont {C.}~\bibnamefont {Wagner}},\ }\href {https://doi.org/10.1002/bbpc.19610650704} {\bibfield  {journal} {\bibinfo  {journal} {Z. Elektrochemie}\ }\textbf {\bibinfo {volume} {65}},\ \bibinfo {pages} {581} (\bibinfo {year} {1961})}\BibitemShut {NoStop}%
\bibitem [{\citenamefont {Binder}\ and\ \citenamefont {Stauffer}(1974)}]{Binder1974}%
  \BibitemOpen
  \bibfield  {author} {\bibinfo {author} {\bibfnamefont {K.}~\bibnamefont {Binder}}\ and\ \bibinfo {author} {\bibfnamefont {D.}~\bibnamefont {Stauffer}},\ }\href {https://doi.org/10.1103/PhysRevLett.33.1006} {\bibfield  {journal} {\bibinfo  {journal} {Phys. Rev. Lett.}\ }\textbf {\bibinfo {volume} {33}},\ \bibinfo {pages} {1006} (\bibinfo {year} {1974})}\BibitemShut {NoStop}%
\bibitem [{\citenamefont {Marro}\ \emph {et~al.}(1979)\citenamefont {Marro}, \citenamefont {Lebowitz},\ and\ \citenamefont {Kalos}}]{Marro1979}%
  \BibitemOpen
  \bibfield  {author} {\bibinfo {author} {\bibfnamefont {J.}~\bibnamefont {Marro}}, \bibinfo {author} {\bibfnamefont {J.~L.}\ \bibnamefont {Lebowitz}},\ and\ \bibinfo {author} {\bibfnamefont {M.~H.}\ \bibnamefont {Kalos}},\ }\href {https://doi.org/10.1103/PhysRevLett.43.282} {\bibfield  {journal} {\bibinfo  {journal} {Phys. Rev. Lett.}\ }\textbf {\bibinfo {volume} {43}},\ \bibinfo {pages} {282} (\bibinfo {year} {1979})}\BibitemShut {NoStop}%
\bibitem [{\citenamefont {Furukawa}(1978)}]{Furukawa1978}%
  \BibitemOpen
  \bibfield  {author} {\bibinfo {author} {\bibfnamefont {H.}~\bibnamefont {Furukawa}},\ }\href {https://doi.org/10.1143/PTP.59.1072} {\bibfield  {journal} {\bibinfo  {journal} {Prog. Theor. Phys.}\ }\textbf {\bibinfo {volume} {59}},\ \bibinfo {pages} {1072} (\bibinfo {year} {1978})}\BibitemShut {NoStop}%
\bibitem [{\citenamefont {Furukawa}(1979)}]{Furukawa1979}%
  \BibitemOpen
  \bibfield  {author} {\bibinfo {author} {\bibfnamefont {H.}~\bibnamefont {Furukawa}},\ }\href {https://doi.org/10.1103/PhysRevLett.43.136} {\bibfield  {journal} {\bibinfo  {journal} {Phys. Rev. Lett.}\ }\textbf {\bibinfo {volume} {43}},\ \bibinfo {pages} {136} (\bibinfo {year} {1979})}\BibitemShut {NoStop}%
\bibitem [{\citenamefont {Furukawa}(1985)}]{Furukawa1985}%
  \BibitemOpen
  \bibfield  {author} {\bibinfo {author} {\bibfnamefont {H.}~\bibnamefont {Furukawa}},\ }\href {https://www.tandfonline.com/doi/abs/10.1080/00018738500101841?utm_source=chatgpt.com} {\bibfield  {journal} {\bibinfo  {journal} {Adv. Phys.}\ } (\bibinfo {year} {1985})}\BibitemShut {NoStop}%
\bibitem [{\citenamefont {Coniglio}\ and\ \citenamefont {Zannetti}(1989)}]{Coniglio1989}%
  \BibitemOpen
  \bibfield  {author} {\bibinfo {author} {\bibfnamefont {A.}~\bibnamefont {Coniglio}}\ and\ \bibinfo {author} {\bibfnamefont {M.}~\bibnamefont {Zannetti}},\ }\href {https://doi.org/10.1209/0295-5075/10/6/012} {\bibfield  {journal} {\bibinfo  {journal} {Europhys. Lett.}\ }\textbf {\bibinfo {volume} {10}},\ \bibinfo {pages} {575} (\bibinfo {year} {1989})}\BibitemShut {NoStop}%
\bibitem [{\citenamefont {Shore}\ and\ \citenamefont {Sethna}(1991)}]{Shore1991}%
  \BibitemOpen
  \bibfield  {author} {\bibinfo {author} {\bibfnamefont {J.~D.}\ \bibnamefont {Shore}}\ and\ \bibinfo {author} {\bibfnamefont {J.~P.}\ \bibnamefont {Sethna}},\ }\href {https://doi.org/10.1103/PhysRevB.43.3782} {\bibfield  {journal} {\bibinfo  {journal} {Phys. Rev. B}\ }\textbf {\bibinfo {volume} {43}},\ \bibinfo {pages} {3782} (\bibinfo {year} {1991})}\BibitemShut {NoStop}%
\bibitem [{\citenamefont {Cugliandolo}(2015)}]{Cugliandolo2015}%
  \BibitemOpen
  \bibfield  {author} {\bibinfo {author} {\bibfnamefont {L.~F.}\ \bibnamefont {Cugliandolo}},\ }\href {https://doi.org/10.1016/j.crhy.2015.02.005} {\bibfield  {journal} {\bibinfo  {journal} {C. R. Phys.}\ }\textbf {\bibinfo {volume} {16}},\ \bibinfo {pages} {257} (\bibinfo {year} {2015})}\BibitemShut {NoStop}%
\bibitem [{\citenamefont {Zechmann}\ \emph {et~al.}(2023)\citenamefont {Zechmann}, \citenamefont {Altman}, \citenamefont {Knap},\ and\ \citenamefont {Feldmeier}}]{Zechmann2023}%
  \BibitemOpen
  \bibfield  {author} {\bibinfo {author} {\bibfnamefont {P.}~\bibnamefont {Zechmann}}, \bibinfo {author} {\bibfnamefont {E.}~\bibnamefont {Altman}}, \bibinfo {author} {\bibfnamefont {M.}~\bibnamefont {Knap}},\ and\ \bibinfo {author} {\bibfnamefont {J.}~\bibnamefont {Feldmeier}},\ }\href {https://doi.org/10.1103/PhysRevB.107.195131} {\bibfield  {journal} {\bibinfo  {journal} {Physical Review B}\ }\textbf {\bibinfo {volume} {107}},\ \bibinfo {pages} {195131} (\bibinfo {year} {2023})}\BibitemShut {NoStop}%
\bibitem [{\citenamefont {Zechmann}\ \emph {et~al.}(2024)\citenamefont {Zechmann}, \citenamefont {Boesl}, \citenamefont {Feldmeier},\ and\ \citenamefont {Knap}}]{Zechmann2024}%
  \BibitemOpen
  \bibfield  {author} {\bibinfo {author} {\bibfnamefont {P.}~\bibnamefont {Zechmann}}, \bibinfo {author} {\bibfnamefont {J.}~\bibnamefont {Boesl}}, \bibinfo {author} {\bibfnamefont {J.}~\bibnamefont {Feldmeier}},\ and\ \bibinfo {author} {\bibfnamefont {M.}~\bibnamefont {Knap}},\ }\href {https://doi.org/10.1103/PhysRevB.109.125137} {\bibfield  {journal} {\bibinfo  {journal} {Phys. Rev. B}\ }\textbf {\bibinfo {volume} {109}},\ \bibinfo {pages} {125137} (\bibinfo {year} {2024})}\BibitemShut {NoStop}%
\bibitem [{\citenamefont {Moudgalya}\ \emph {et~al.}(2020)\citenamefont {Moudgalya}, \citenamefont {Prem}, \citenamefont {Nandkishore}, \citenamefont {Regnault},\ and\ \citenamefont {Bernevig}}]{Moudgalya2020}%
  \BibitemOpen
  \bibfield  {author} {\bibinfo {author} {\bibfnamefont {S.}~\bibnamefont {Moudgalya}}, \bibinfo {author} {\bibfnamefont {A.}~\bibnamefont {Prem}}, \bibinfo {author} {\bibfnamefont {R.}~\bibnamefont {Nandkishore}}, \bibinfo {author} {\bibfnamefont {N.}~\bibnamefont {Regnault}},\ and\ \bibinfo {author} {\bibfnamefont {B.~A.}\ \bibnamefont {Bernevig}},\ }in\ \href {https://doi.org/10.1142/9789811231711_0009} {\emph {\bibinfo {booktitle} {{Memorial Volume for Shoucheng Zhang}}}}\ (\bibinfo  {publisher} {World Scientific},\ \bibinfo {address} {Singapore},\ \bibinfo {year} {2020})\ pp.\ \bibinfo {pages} {147--209}\BibitemShut {NoStop}%
\bibitem [{\citenamefont {Dubinkin}\ \emph {et~al.}(2021)\citenamefont {Dubinkin}, \citenamefont {{May-Mann}},\ and\ \citenamefont {Hughes}}]{Dubinkin2021}%
  \BibitemOpen
  \bibfield  {author} {\bibinfo {author} {\bibfnamefont {O.}~\bibnamefont {Dubinkin}}, \bibinfo {author} {\bibfnamefont {J.}~\bibnamefont {{May-Mann}}},\ and\ \bibinfo {author} {\bibfnamefont {T.~L.}\ \bibnamefont {Hughes}},\ }\href {https://doi.org/10.1103/PhysRevB.103.125129} {\bibfield  {journal} {\bibinfo  {journal} {Physical Review B}\ }\textbf {\bibinfo {volume} {103}},\ \bibinfo {pages} {125129} (\bibinfo {year} {2021})},\ \Eprint {https://arxiv.org/abs/1909.07403} {arxiv:1909.07403 [cond-mat]} \BibitemShut {NoStop}%
\bibitem [{\citenamefont {Stahl}\ \emph {et~al.}(2022)\citenamefont {Stahl}, \citenamefont {Lake},\ and\ \citenamefont {Nandkishore}}]{Stahl2022}%
  \BibitemOpen
  \bibfield  {author} {\bibinfo {author} {\bibfnamefont {C.}~\bibnamefont {Stahl}}, \bibinfo {author} {\bibfnamefont {E.}~\bibnamefont {Lake}},\ and\ \bibinfo {author} {\bibfnamefont {R.}~\bibnamefont {Nandkishore}},\ }\href {https://doi.org/10.1103/PhysRevB.105.155107} {\bibfield  {journal} {\bibinfo  {journal} {Physical Review B}\ }\textbf {\bibinfo {volume} {105}},\ \bibinfo {pages} {155107} (\bibinfo {year} {2022})}\BibitemShut {NoStop}%
\bibitem [{\citenamefont {Morningstar}\ \emph {et~al.}(2020)\citenamefont {Morningstar}, \citenamefont {Khemani},\ and\ \citenamefont {Huse}}]{Morningstar2020}%
  \BibitemOpen
  \bibfield  {author} {\bibinfo {author} {\bibfnamefont {A.}~\bibnamefont {Morningstar}}, \bibinfo {author} {\bibfnamefont {V.}~\bibnamefont {Khemani}},\ and\ \bibinfo {author} {\bibfnamefont {D.~A.}\ \bibnamefont {Huse}},\ }\href {https://doi.org/10.1103/PhysRevB.101.214205} {\bibfield  {journal} {\bibinfo  {journal} {Physical Review B}\ }\textbf {\bibinfo {volume} {101}},\ \bibinfo {pages} {214205} (\bibinfo {year} {2020})}\BibitemShut {NoStop}%
\bibitem [{\citenamefont {Pozderac}\ \emph {et~al.}(2023)\citenamefont {Pozderac}, \citenamefont {Speck}, \citenamefont {Feng}, \citenamefont {Huse},\ and\ \citenamefont {Skinner}}]{Pozderac_2023}%
  \BibitemOpen
  \bibfield  {author} {\bibinfo {author} {\bibfnamefont {C.}~\bibnamefont {Pozderac}}, \bibinfo {author} {\bibfnamefont {S.}~\bibnamefont {Speck}}, \bibinfo {author} {\bibfnamefont {X.}~\bibnamefont {Feng}}, \bibinfo {author} {\bibfnamefont {D.~A.}\ \bibnamefont {Huse}},\ and\ \bibinfo {author} {\bibfnamefont {B.}~\bibnamefont {Skinner}},\ }\href {https://doi.org/10.1103/PhysRevB.107.045137} {\bibfield  {journal} {\bibinfo  {journal} {Phys. Rev. B}\ }\textbf {\bibinfo {volume} {107}},\ \bibinfo {pages} {045137} (\bibinfo {year} {2023})}\BibitemShut {NoStop}%
\bibitem [{\citenamefont {Patil}\ \emph {et~al.}(2023)\citenamefont {Patil}, \citenamefont {Heyl},\ and\ \citenamefont {Alet}}]{Patil2023}%
  \BibitemOpen
  \bibfield  {author} {\bibinfo {author} {\bibfnamefont {P.}~\bibnamefont {Patil}}, \bibinfo {author} {\bibfnamefont {M.}~\bibnamefont {Heyl}},\ and\ \bibinfo {author} {\bibfnamefont {F.}~\bibnamefont {Alet}},\ }\href {https://doi.org/10.1103/PhysRevE.107.034119} {\bibfield  {journal} {\bibinfo  {journal} {Phys. Rev. E}\ }\textbf {\bibinfo {volume} {107}},\ \bibinfo {pages} {034119} (\bibinfo {year} {2023})}\BibitemShut {NoStop}%
\bibitem [{\citenamefont {Burnell}\ \emph {et~al.}(2024)\citenamefont {Burnell}, \citenamefont {Moudgalya},\ and\ \citenamefont {Prem}}]{Burnell_2024}%
  \BibitemOpen
  \bibfield  {author} {\bibinfo {author} {\bibfnamefont {F.~J.}\ \bibnamefont {Burnell}}, \bibinfo {author} {\bibfnamefont {S.}~\bibnamefont {Moudgalya}},\ and\ \bibinfo {author} {\bibfnamefont {A.}~\bibnamefont {Prem}},\ }\href {https://doi.org/10.1103/PhysRevB.110.L121113} {\bibfield  {journal} {\bibinfo  {journal} {Phys. Rev. B}\ }\textbf {\bibinfo {volume} {110}},\ \bibinfo {pages} {L121113} (\bibinfo {year} {2024})}\BibitemShut {NoStop}%
\bibitem [{\citenamefont {Gliozzi}\ \emph {et~al.}(2024)\citenamefont {Gliozzi}, \citenamefont {De~Tomasi},\ and\ \citenamefont {Hughes}}]{Gliozzi_2024}%
  \BibitemOpen
  \bibfield  {author} {\bibinfo {author} {\bibfnamefont {J.}~\bibnamefont {Gliozzi}}, \bibinfo {author} {\bibfnamefont {G.}~\bibnamefont {De~Tomasi}},\ and\ \bibinfo {author} {\bibfnamefont {T.~L.}\ \bibnamefont {Hughes}},\ }\href {https://doi.org/10.1103/PhysRevLett.133.136503} {\bibfield  {journal} {\bibinfo  {journal} {Phys. Rev. Lett.}\ }\textbf {\bibinfo {volume} {133}},\ \bibinfo {pages} {136503} (\bibinfo {year} {2024})}\BibitemShut {NoStop}%
\bibitem [{\citenamefont {Gliozzi}\ \emph {et~al.}(2025)\citenamefont {Gliozzi}, \citenamefont {Balducci}, \citenamefont {Hughes},\ and\ \citenamefont {De~Tomasi}}]{Gliozzi2025}%
  \BibitemOpen
  \bibfield  {author} {\bibinfo {author} {\bibfnamefont {J.}~\bibnamefont {Gliozzi}}, \bibinfo {author} {\bibfnamefont {F.}~\bibnamefont {Balducci}}, \bibinfo {author} {\bibfnamefont {T.~L.}\ \bibnamefont {Hughes}},\ and\ \bibinfo {author} {\bibfnamefont {G.}~\bibnamefont {De~Tomasi}},\ }\href@noop {} {\  (\bibinfo {year} {2025})},\ \Eprint {https://arxiv.org/abs/2504.10580} {arXiv:2504.10580} \BibitemShut {NoStop}%
\bibitem [{\citenamefont {De~Tomasi}\ \emph {et~al.}(2019)\citenamefont {De~Tomasi}, \citenamefont {Hetterich}, \citenamefont {Sala},\ and\ \citenamefont {Pollmann}}]{DeTomasi2019}%
  \BibitemOpen
  \bibfield  {author} {\bibinfo {author} {\bibfnamefont {G.}~\bibnamefont {De~Tomasi}}, \bibinfo {author} {\bibfnamefont {D.}~\bibnamefont {Hetterich}}, \bibinfo {author} {\bibfnamefont {P.}~\bibnamefont {Sala}},\ and\ \bibinfo {author} {\bibfnamefont {F.}~\bibnamefont {Pollmann}},\ }\href {https://doi.org/10.1103/PhysRevB.100.214313} {\bibfield  {journal} {\bibinfo  {journal} {Phys. Rev. B}\ }\textbf {\bibinfo {volume} {100}},\ \bibinfo {pages} {214313} (\bibinfo {year} {2019})}\BibitemShut {NoStop}%
\bibitem [{\citenamefont {Yang}\ \emph {et~al.}(2020)\citenamefont {Yang}, \citenamefont {Liu}, \citenamefont {Gorshkov},\ and\ \citenamefont {Iadecola}}]{Yang2020}%
  \BibitemOpen
  \bibfield  {author} {\bibinfo {author} {\bibfnamefont {Z.-C.}\ \bibnamefont {Yang}}, \bibinfo {author} {\bibfnamefont {F.}~\bibnamefont {Liu}}, \bibinfo {author} {\bibfnamefont {A.~V.}\ \bibnamefont {Gorshkov}},\ and\ \bibinfo {author} {\bibfnamefont {T.}~\bibnamefont {Iadecola}},\ }\href {https://doi.org/10.1103/PhysRevLett.124.207602} {\bibfield  {journal} {\bibinfo  {journal} {Phys. Rev. Lett.}\ }\textbf {\bibinfo {volume} {124}},\ \bibinfo {pages} {207602} (\bibinfo {year} {2020})}\BibitemShut {NoStop}%
\bibitem [{\citenamefont {Cornell}\ \emph {et~al.}(1991)\citenamefont {Cornell}, \citenamefont {Kaski},\ and\ \citenamefont {Stinchcombe}}]{Cornell1991}%
  \BibitemOpen
  \bibfield  {author} {\bibinfo {author} {\bibfnamefont {S.~J.}\ \bibnamefont {Cornell}}, \bibinfo {author} {\bibfnamefont {K.}~\bibnamefont {Kaski}},\ and\ \bibinfo {author} {\bibfnamefont {R.~B.}\ \bibnamefont {Stinchcombe}},\ }\href {https://doi.org/10.1103/PhysRevB.44.12263} {\bibfield  {journal} {\bibinfo  {journal} {Phys. Rev. B}\ }\textbf {\bibinfo {volume} {44}},\ \bibinfo {pages} {12263} (\bibinfo {year} {1991})}\BibitemShut {NoStop}%
\bibitem [{\citenamefont {Ben-Naim}\ and\ \citenamefont {Krapivsky}(1998)}]{Ben-Naim1998}%
  \BibitemOpen
  \bibfield  {author} {\bibinfo {author} {\bibfnamefont {E.}~\bibnamefont {Ben-Naim}}\ and\ \bibinfo {author} {\bibfnamefont {P.~L.}\ \bibnamefont {Krapivsky}},\ }\href {https://doi.org/10.1023/B:JOSS.0000033243.27556.99} {\bibfield  {journal} {\bibinfo  {journal} {J. Stat. Phys.}\ }\textbf {\bibinfo {volume} {93}},\ \bibinfo {pages} {583} (\bibinfo {year} {1998})}\BibitemShut {NoStop}%
\bibitem [{\citenamefont {Cardy}(1996)}]{Cardy}%
  \BibitemOpen
  \bibfield  {author} {\bibinfo {author} {\bibfnamefont {J.}~\bibnamefont {Cardy}},\ }\href@noop {} {\emph {\bibinfo {title} {{Scaling and Renormalization in Statistical Physics}}}}\ (\bibinfo  {publisher} {Cambridge University Press},\ \bibinfo {year} {1996})\BibitemShut {NoStop}%
\bibitem [{\citenamefont {Hohenberg}\ and\ \citenamefont {Halperin}(1977)}]{Hohenberg1977}%
  \BibitemOpen
  \bibfield  {author} {\bibinfo {author} {\bibfnamefont {P.~C.}\ \bibnamefont {Hohenberg}}\ and\ \bibinfo {author} {\bibfnamefont {B.~I.}\ \bibnamefont {Halperin}},\ }\href {https://doi.org/10.1103/RevModPhys.49.435} {\bibfield  {journal} {\bibinfo  {journal} {Rev. Mod. Phys.}\ }\textbf {\bibinfo {volume} {49}},\ \bibinfo {pages} {435} (\bibinfo {year} {1977})}\BibitemShut {NoStop}%
\bibitem [{\citenamefont {Cahn}\ and\ \citenamefont {Hilliard}(1958)}]{Cahn1958}%
  \BibitemOpen
  \bibfield  {author} {\bibinfo {author} {\bibfnamefont {J.~W.}\ \bibnamefont {Cahn}}\ and\ \bibinfo {author} {\bibfnamefont {J.~E.}\ \bibnamefont {Hilliard}},\ }\href {https://doi.org/10.1063/1.1744102} {\bibfield  {journal} {\bibinfo  {journal} {J. Chem. Phys.}\ }\textbf {\bibinfo {volume} {28}},\ \bibinfo {pages} {258} (\bibinfo {year} {1958})}\BibitemShut {NoStop}%
\bibitem [{\citenamefont {Huse}(1986)}]{Huse1986}%
  \BibitemOpen
  \bibfield  {author} {\bibinfo {author} {\bibfnamefont {D.~A.}\ \bibnamefont {Huse}},\ }\href {https://doi.org/10.1103/PhysRevB.34.7845} {\bibfield  {journal} {\bibinfo  {journal} {Phys. Rev. B}\ }\textbf {\bibinfo {volume} {34}},\ \bibinfo {pages} {7845} (\bibinfo {year} {1986})}\BibitemShut {NoStop}%
\bibitem [{\citenamefont {Allen}\ and\ \citenamefont {Cahn}(1972)}]{Allen1972}%
  \BibitemOpen
  \bibfield  {author} {\bibinfo {author} {\bibfnamefont {S.~M.}\ \bibnamefont {Allen}}\ and\ \bibinfo {author} {\bibfnamefont {J.~W.}\ \bibnamefont {Cahn}},\ }\href {https://doi.org/10.1016/0001-6160(72)90037-5} {\bibfield  {journal} {\bibinfo  {journal} {Acta Metall.}\ }\textbf {\bibinfo {volume} {20}},\ \bibinfo {pages} {423} (\bibinfo {year} {1972})}\BibitemShut {NoStop}%
\bibitem [{\citenamefont {Bray}(1998)}]{Bray1998perturb}%
  \BibitemOpen
  \bibfield  {author} {\bibinfo {author} {\bibfnamefont {A.~J.}\ \bibnamefont {Bray}},\ }\href {https://doi.org/10.1103/PhysRevE.58.1508} {\bibfield  {journal} {\bibinfo  {journal} {Phys. Rev. E}\ }\textbf {\bibinfo {volume} {58}},\ \bibinfo {pages} {1508} (\bibinfo {year} {1998})}\BibitemShut {NoStop}%
\bibitem [{\citenamefont {van Gemmert}\ \emph {et~al.}(2005)\citenamefont {van Gemmert}, \citenamefont {Barkema},\ and\ \citenamefont {Puri}}]{vanGemmert2005}%
  \BibitemOpen
  \bibfield  {author} {\bibinfo {author} {\bibfnamefont {S.}~\bibnamefont {van Gemmert}}, \bibinfo {author} {\bibfnamefont {G.~T.}\ \bibnamefont {Barkema}},\ and\ \bibinfo {author} {\bibfnamefont {S.}~\bibnamefont {Puri}},\ }\href {https://doi.org/10.1103/PhysRevE.72.046131} {\bibfield  {journal} {\bibinfo  {journal} {Phys. Rev. E}\ }\textbf {\bibinfo {volume} {72}},\ \bibinfo {pages} {046131} (\bibinfo {year} {2005})}\BibitemShut {NoStop}%
\bibitem [{\citenamefont {Amar}\ \emph {et~al.}(1988)\citenamefont {Amar}, \citenamefont {Sullivan},\ and\ \citenamefont {Mountain}}]{Amar1988}%
  \BibitemOpen
  \bibfield  {author} {\bibinfo {author} {\bibfnamefont {J.~G.}\ \bibnamefont {Amar}}, \bibinfo {author} {\bibfnamefont {F.~E.}\ \bibnamefont {Sullivan}},\ and\ \bibinfo {author} {\bibfnamefont {R.~D.}\ \bibnamefont {Mountain}},\ }\href {https://doi.org/10.1103/PhysRevB.37.196} {\bibfield  {journal} {\bibinfo  {journal} {Phys. Rev. B}\ }\textbf {\bibinfo {volume} {37}},\ \bibinfo {pages} {196} (\bibinfo {year} {1988})}\BibitemShut {NoStop}%
\bibitem [{\citenamefont {Sala}\ \emph {et~al.}(2022)\citenamefont {Sala}, \citenamefont {Lehmann}, \citenamefont {Rakovszky},\ and\ \citenamefont {Pollmann}}]{Sala2022}%
  \BibitemOpen
  \bibfield  {author} {\bibinfo {author} {\bibfnamefont {P.}~\bibnamefont {Sala}}, \bibinfo {author} {\bibfnamefont {J.}~\bibnamefont {Lehmann}}, \bibinfo {author} {\bibfnamefont {T.}~\bibnamefont {Rakovszky}},\ and\ \bibinfo {author} {\bibfnamefont {F.}~\bibnamefont {Pollmann}},\ }\href {https://doi.org/10.1103/PhysRevLett.129.170601} {\bibfield  {journal} {\bibinfo  {journal} {Phys. Rev. Lett.}\ }\textbf {\bibinfo {volume} {129}},\ \bibinfo {pages} {170601} (\bibinfo {year} {2022})}\BibitemShut {NoStop}%
\bibitem [{\citenamefont {Chandran}\ \emph {et~al.}(2023)\citenamefont {Chandran}, \citenamefont {Iadecola}, \citenamefont {Khemani},\ and\ \citenamefont {Moessner}}]{Chandran2023}%
  \BibitemOpen
  \bibfield  {author} {\bibinfo {author} {\bibfnamefont {A.}~\bibnamefont {Chandran}}, \bibinfo {author} {\bibfnamefont {T.}~\bibnamefont {Iadecola}}, \bibinfo {author} {\bibfnamefont {V.}~\bibnamefont {Khemani}},\ and\ \bibinfo {author} {\bibfnamefont {R.}~\bibnamefont {Moessner}},\ }\href {https://doi.org/10.1146/annurev-conmatphys-031620-101617} {\bibfield  {journal} {\bibinfo  {journal} {Annu. Rev. Condens. Matter Phys.}\ ,\ \bibinfo {pages} {443}} (\bibinfo {year} {2023})}\BibitemShut {NoStop}%
\bibitem [{\citenamefont {Flajolet}\ and\ \citenamefont {Sedgewick}(2009)}]{flajolet}%
  \BibitemOpen
  \bibfield  {author} {\bibinfo {author} {\bibfnamefont {P.}~\bibnamefont {Flajolet}}\ and\ \bibinfo {author} {\bibfnamefont {R.}~\bibnamefont {Sedgewick}},\ }\href@noop {} {\emph {\bibinfo {title} {Analytic Combinatorics}}}\ (\bibinfo  {publisher} {Cambridge University Press},\ \bibinfo {year} {2009})\BibitemShut {NoStop}%
\bibitem [{\citenamefont {Sollich}\ and\ \citenamefont {Evans}(1999)}]{Sollich1999}%
  \BibitemOpen
  \bibfield  {author} {\bibinfo {author} {\bibfnamefont {P.}~\bibnamefont {Sollich}}\ and\ \bibinfo {author} {\bibfnamefont {M.~R.}\ \bibnamefont {Evans}},\ }\href {https://doi.org/10.1103/PhysRevLett.83.3238} {\bibfield  {journal} {\bibinfo  {journal} {Phys. Rev. Lett.}\ }\textbf {\bibinfo {volume} {83}},\ \bibinfo {pages} {3238} (\bibinfo {year} {1999})}\BibitemShut {NoStop}%
\bibitem [{\citenamefont {Sollich}\ and\ \citenamefont {Evans}(2003)}]{Sollich2003}%
  \BibitemOpen
  \bibfield  {author} {\bibinfo {author} {\bibfnamefont {P.}~\bibnamefont {Sollich}}\ and\ \bibinfo {author} {\bibfnamefont {M.~R.}\ \bibnamefont {Evans}},\ }\href {https://doi.org/10.1103/PhysRevE.68.031504} {\bibfield  {journal} {\bibinfo  {journal} {Phys. Rev. E}\ }\textbf {\bibinfo {volume} {68}},\ \bibinfo {pages} {031504} (\bibinfo {year} {2003})}\BibitemShut {NoStop}%
\bibitem [{\citenamefont {Placke}\ \emph {et~al.}(2024)\citenamefont {Placke}, \citenamefont {Benton},\ and\ \citenamefont {Moessner}}]{Placke2024Jul}%
  \BibitemOpen
  \bibfield  {author} {\bibinfo {author} {\bibfnamefont {B.}~\bibnamefont {Placke}}, \bibinfo {author} {\bibfnamefont {O.}~\bibnamefont {Benton}},\ and\ \bibinfo {author} {\bibfnamefont {R.}~\bibnamefont {Moessner}},\ }\href {https://doi.org/10.1103/PhysRevB.110.L020401} {\bibfield  {journal} {\bibinfo  {journal} {Phys. Rev. B}\ }\textbf {\bibinfo {volume} {110}},\ \bibinfo {pages} {L020401} (\bibinfo {year} {2024})}\BibitemShut {NoStop}%
\bibitem [{\citenamefont {Turkowski}\ \emph {et~al.}(2006)\citenamefont {Turkowski}, \citenamefont {Sacramento},\ and\ \citenamefont {Vieira}}]{Turkowski2006}%
  \BibitemOpen
  \bibfield  {author} {\bibinfo {author} {\bibfnamefont {V.~M.}\ \bibnamefont {Turkowski}}, \bibinfo {author} {\bibfnamefont {P.~D.}\ \bibnamefont {Sacramento}},\ and\ \bibinfo {author} {\bibfnamefont {V.~R.}\ \bibnamefont {Vieira}},\ }\href {https://doi.org/10.1103/PhysRevB.73.214437} {\bibfield  {journal} {\bibinfo  {journal} {Phys. Rev. B}\ }\textbf {\bibinfo {volume} {73}},\ \bibinfo {pages} {214437} (\bibinfo {year} {2006})}\BibitemShut {NoStop}%
\bibitem [{\citenamefont {Aron}\ \emph {et~al.}(2009)\citenamefont {Aron}, \citenamefont {Biroli},\ and\ \citenamefont {Cugliandolo}}]{Aron2009}%
  \BibitemOpen
  \bibfield  {author} {\bibinfo {author} {\bibfnamefont {C.}~\bibnamefont {Aron}}, \bibinfo {author} {\bibfnamefont {G.}~\bibnamefont {Biroli}},\ and\ \bibinfo {author} {\bibfnamefont {L.~F.}\ \bibnamefont {Cugliandolo}},\ }\href {https://doi.org/10.1103/PhysRevLett.102.050404} {\bibfield  {journal} {\bibinfo  {journal} {Phys. Rev. Lett.}\ }\textbf {\bibinfo {volume} {102}},\ \bibinfo {pages} {050404} (\bibinfo {year} {2009})}\BibitemShut {NoStop}%
\bibitem [{\citenamefont {Chandran}\ \emph {et~al.}(2013)\citenamefont {Chandran}, \citenamefont {Nanduri}, \citenamefont {Gubser},\ and\ \citenamefont {Sondhi}}]{Chandran2013}%
  \BibitemOpen
  \bibfield  {author} {\bibinfo {author} {\bibfnamefont {A.}~\bibnamefont {Chandran}}, \bibinfo {author} {\bibfnamefont {A.}~\bibnamefont {Nanduri}}, \bibinfo {author} {\bibfnamefont {S.~S.}\ \bibnamefont {Gubser}},\ and\ \bibinfo {author} {\bibfnamefont {S.~L.}\ \bibnamefont {Sondhi}},\ }\href {https://doi.org/10.1103/PhysRevB.88.024306} {\bibfield  {journal} {\bibinfo  {journal} {Phys. Rev. B}\ }\textbf {\bibinfo {volume} {88}},\ \bibinfo {pages} {024306} (\bibinfo {year} {2013})}\BibitemShut {NoStop}%
\bibitem [{\citenamefont {Maraga}\ \emph {et~al.}(2015)\citenamefont {Maraga}, \citenamefont {Chiocchetta}, \citenamefont {Mitra},\ and\ \citenamefont {Gambassi}}]{Maraga2015}%
  \BibitemOpen
  \bibfield  {author} {\bibinfo {author} {\bibfnamefont {A.}~\bibnamefont {Maraga}}, \bibinfo {author} {\bibfnamefont {A.}~\bibnamefont {Chiocchetta}}, \bibinfo {author} {\bibfnamefont {A.}~\bibnamefont {Mitra}},\ and\ \bibinfo {author} {\bibfnamefont {A.}~\bibnamefont {Gambassi}},\ }\href {https://doi.org/10.1103/PhysRevE.92.042151} {\bibfield  {journal} {\bibinfo  {journal} {Phys. Rev. E}\ }\textbf {\bibinfo {volume} {92}},\ \bibinfo {pages} {042151} (\bibinfo {year} {2015})}\BibitemShut {NoStop}%
\bibitem [{\citenamefont {Samajdar}\ and\ \citenamefont {Huse}(2024)}]{Samajdar2024}%
  \BibitemOpen
  \bibfield  {author} {\bibinfo {author} {\bibfnamefont {R.}~\bibnamefont {Samajdar}}\ and\ \bibinfo {author} {\bibfnamefont {D.~A.}\ \bibnamefont {Huse}},\ }\href@noop {} {\  (\bibinfo {year} {2024})},\ \Eprint {https://arxiv.org/abs/2401.15144} {2401.15144} \BibitemShut {NoStop}%
\bibitem [{\citenamefont {Balducci}\ \emph {et~al.}(2025)\citenamefont {Balducci}, \citenamefont {Chandran},\ and\ \citenamefont {Moessner}}]{Balducci2025}%
  \BibitemOpen
  \bibfield  {author} {\bibinfo {author} {\bibfnamefont {F.}~\bibnamefont {Balducci}}, \bibinfo {author} {\bibfnamefont {A.}~\bibnamefont {Chandran}},\ and\ \bibinfo {author} {\bibfnamefont {R.}~\bibnamefont {Moessner}},\ }\href@noop {} {\  (\bibinfo {year} {2025})},\ \Eprint {https://arxiv.org/abs/2507.17386} {2507.17386} \BibitemShut {NoStop}%
\bibitem [{\citenamefont {Andersen}\ \emph {et~al.}(2025)\citenamefont {Andersen}, \citenamefont {Astrakhantsev}, \citenamefont {Karamlou} \emph {et~al.}}]{Andersen2025Thermalization}%
  \BibitemOpen
  \bibfield  {author} {\bibinfo {author} {\bibfnamefont {T.~I.}\ \bibnamefont {Andersen}}, \bibinfo {author} {\bibfnamefont {N.}~\bibnamefont {Astrakhantsev}}, \bibinfo {author} {\bibfnamefont {A.~H.}\ \bibnamefont {Karamlou}}, \emph {et~al.},\ }\href {https://doi.org/10.1038/s41586-024-08460-3} {\bibfield  {journal} {\bibinfo  {journal} {Nature}\ }\textbf {\bibinfo {volume} {638}},\ \bibinfo {pages} {79} (\bibinfo {year} {2025})}\BibitemShut {NoStop}%
\bibitem [{\citenamefont {Manovitz}\ \emph {et~al.}(2025)\citenamefont {Manovitz}, \citenamefont {Li}, \citenamefont {Ebadi} \emph {et~al.}}]{Manovitz2025Quantum}%
  \BibitemOpen
  \bibfield  {author} {\bibinfo {author} {\bibfnamefont {T.}~\bibnamefont {Manovitz}}, \bibinfo {author} {\bibfnamefont {S.~H.}\ \bibnamefont {Li}}, \bibinfo {author} {\bibfnamefont {S.}~\bibnamefont {Ebadi}}, \emph {et~al.},\ }\href {https://doi.org/10.1038/s41586-024-08353-5} {\bibfield  {journal} {\bibinfo  {journal} {Nature}\ }\textbf {\bibinfo {volume} {638}},\ \bibinfo {pages} {86} (\bibinfo {year} {2025})}\BibitemShut {NoStop}%
\bibitem [{\citenamefont {Guardado-Sanchez}\ \emph {et~al.}(2020)\citenamefont {Guardado-Sanchez}, \citenamefont {Morningstar}, \citenamefont {Spar}, \citenamefont {Brown}, \citenamefont {Huse},\ and\ \citenamefont {Bakr}}]{Guardado-Sanchez2020}%
  \BibitemOpen
  \bibfield  {author} {\bibinfo {author} {\bibfnamefont {E.}~\bibnamefont {Guardado-Sanchez}}, \bibinfo {author} {\bibfnamefont {A.}~\bibnamefont {Morningstar}}, \bibinfo {author} {\bibfnamefont {B.~M.}\ \bibnamefont {Spar}}, \bibinfo {author} {\bibfnamefont {P.~T.}\ \bibnamefont {Brown}}, \bibinfo {author} {\bibfnamefont {D.~A.}\ \bibnamefont {Huse}},\ and\ \bibinfo {author} {\bibfnamefont {W.~S.}\ \bibnamefont {Bakr}},\ }\href {https://doi.org/10.1103/PhysRevX.10.011042} {\bibfield  {journal} {\bibinfo  {journal} {Phys. Rev. X}\ }\textbf {\bibinfo {volume} {10}},\ \bibinfo {pages} {011042} (\bibinfo {year} {2020})}\BibitemShut {NoStop}%
\bibitem [{\citenamefont {Scherg}\ \emph {et~al.}(2021)\citenamefont {Scherg}, \citenamefont {Kohlert}, \citenamefont {Sala}, \citenamefont {Pollmann}, \citenamefont {Hebbe~Madhusudhana}, \citenamefont {Bloch},\ and\ \citenamefont {Aidelsburger}}]{Scherg2021}%
  \BibitemOpen
  \bibfield  {author} {\bibinfo {author} {\bibfnamefont {S.}~\bibnamefont {Scherg}}, \bibinfo {author} {\bibfnamefont {T.}~\bibnamefont {Kohlert}}, \bibinfo {author} {\bibfnamefont {P.}~\bibnamefont {Sala}}, \bibinfo {author} {\bibfnamefont {F.}~\bibnamefont {Pollmann}}, \bibinfo {author} {\bibfnamefont {B.}~\bibnamefont {Hebbe~Madhusudhana}}, \bibinfo {author} {\bibfnamefont {I.}~\bibnamefont {Bloch}},\ and\ \bibinfo {author} {\bibfnamefont {M.}~\bibnamefont {Aidelsburger}},\ }\href {https://doi.org/10.1038/s41467-021-24726-0} {\bibfield  {journal} {\bibinfo  {journal} {Nat. Commun.}\ }\textbf {\bibinfo {volume} {12}},\ \bibinfo {pages} {1} (\bibinfo {year} {2021})}\BibitemShut {NoStop}%
\bibitem [{\citenamefont {Kohlert}\ \emph {et~al.}(2023)\citenamefont {Kohlert}, \citenamefont {Scherg}, \citenamefont {Sala}, \citenamefont {Pollmann}, \citenamefont {Hebbe~Madhusudhana}, \citenamefont {Bloch},\ and\ \citenamefont {Aidelsburger}}]{Kohlert2023}%
  \BibitemOpen
  \bibfield  {author} {\bibinfo {author} {\bibfnamefont {T.}~\bibnamefont {Kohlert}}, \bibinfo {author} {\bibfnamefont {S.}~\bibnamefont {Scherg}}, \bibinfo {author} {\bibfnamefont {P.}~\bibnamefont {Sala}}, \bibinfo {author} {\bibfnamefont {F.}~\bibnamefont {Pollmann}}, \bibinfo {author} {\bibfnamefont {B.}~\bibnamefont {Hebbe~Madhusudhana}}, \bibinfo {author} {\bibfnamefont {I.}~\bibnamefont {Bloch}},\ and\ \bibinfo {author} {\bibfnamefont {M.}~\bibnamefont {Aidelsburger}},\ }\href {https://doi.org/10.1103/PhysRevLett.130.010201} {\bibfield  {journal} {\bibinfo  {journal} {Phys. Rev. Lett.}\ }\textbf {\bibinfo {volume} {130}},\ \bibinfo {pages} {010201} (\bibinfo {year} {2023})}\BibitemShut {NoStop}%
\end{thebibliography}%

\end{document}